\documentclass{aa}  
\usepackage{xcolor}

\usepackage[breaklinks=true]{hyperref}
\usepackage{natbib,twoopt}
\bibpunct{(}{)}{;}{a}{}{,}             %% natbib format for A&A and ApJ
\makeatletter
  \newcommandtwoopt{\citeads}[3][][]{\href{http://adsabs.harvard.edu/abs/#3}%
    {\def\hyper@linkstart##1##2{}%
     \let\hyper@linkend\@empty\citealp[#1][#2]{#3}}}
  \newcommandtwoopt{\citepads}[3][][]{\href{http://adsabs.harvard.edu/abs/#3}%
    {\def\hyper@linkstart##1##2{}%
     \let\hyper@linkend\@empty\citep[#1][#2]{#3}}}
  \newcommandtwoopt{\citetads}[3][][]{\href{http://adsabs.harvard.edu/abs/#3}%
    {\def\hyper@linkstart##1##2{}%
     \let\hyper@linkend\@empty\citet[#1][#2]{#3}}}
  \newcommandtwoopt{\citeyearads}[3][][]%
    {\href{http://adsabs.harvard.edu/abs/#3}
    {\def\hyper@linkstart##1##2{}%
     \let\hyper@linkend\@empty\citeyear[#1][#2]{#3}}}
\makeatother
\usepackage{graphicx}
\usepackage{booktabs}
\usepackage{lipsum}
\usepackage{txfonts}
\usepackage{hyperref}
\usepackage{tabularx}
\usepackage{upgreek}

\graphicspath{{Figures/}} %Setting the graphicspath

\begin{document}

   \title{Winds of change: the nuclear and galaxy-scale outflows and the X-ray variability of 2MASS 0918+2117}

   \author{P. Baldini
          \inst{1,2,3}  
    \and G. Lanzuisi\inst{2}  
    \and M. Brusa\inst{1,2}  
    \and A. Merloni\inst{3}  
    \and K. Gkimisi\inst{1}
    \and M. Perna\inst{4}
    \and I.E. López\inst{1,2}
    \and E. Bertola\inst{5,1,2} 
    \and Z. Igo\inst{3,6}  
    \and S. Waddell\inst{3} 
    \and B. Musiimenta\inst{1,2}
    \and C. Aydar\inst{3}
    \and R. Arcodia\thanks{NASA Einstein fellow} \inst{7,3}
    \and G. \textcolor{black}{A.} Matzeu\inst{1,2,8}  
    \and A. Luminari\inst{9,10}
    \and J. Buchner\inst{3}
    \and C. Vignali\inst{1,2}
    \and M. Dadina\inst{2}
    \and A. Comastri\inst{2}
    \and G. Cresci\inst{5}
    \and S. Marchesi\inst{1,2,11}
    \and R. Gilli\inst{2}
    \and F. Tombesi\inst{12,13,14}
    \and \textcolor{black}{R. Serafinelli}\inst{9}
%\and Serafinelli 
} 

   \institute{Dipartimento di Fisica e Astronomia “Augusto Righi”, Università 
              di Bologna, via Gobetti 93/2, 40129 Bologna, Italy \\
              e-mail: pietro.baldini@studio.unibo.it
         \and
              INAF – Osservatorio di Astrofisica e Scienza dello Spazio di Bologna, via Gobetti 93/3, 40129 Bologna, Italy
        \and Max-Planck-Institut f\"ur extraterrestrische Physik, Giessenbachstra{\ss}e 1, D-85748 Garching bei M\"unchen, Germany
        \and 
             Centro de Astrobiolog\'{\i}a (CAB), CSIC-INTA, Ctra. de Ajalvir km 4, Torrej\'on de Ardoz, E-28850, Madrid, Spain.
        %\and European Space Agency, c/o STScI, 3700 San Martin Drive, Baltimore, MD 21218, USA.
        \and    INAF -- Osservatorio Astrofisico di Arcetri, Largo Enrico Fermi 5, I-50125 Firenze, Italy
        \and Exzellenzcluster ORIGINS, Boltzmannstr. 2, 85748, Garching, Germany
         \and MIT Kavli Institute for Astrophysics and Space Research, 70 Vassar Street, Cambridge, MA 02139, USA.
        \and \textcolor{black}{Quasar Science Resources SL for ESA, European Space Astronomy Centre (ESAC), Science Operations Department, 28692, Villanueva de la Ca\~{n}ada, Madrid, Spain}
       \and INAF—Istituto di Astrofisica e Planetologia Spaziali, Via del Fosso del Caveliere 100, I-00133 Roma, Italy. 
        \and INAF—Osservatorio Astronomico di Roma, Via Frascati 33, I-00078 Monte Porzio Catone Roma, Italy.
        \and Department of Physics and Astronomy, Clemson University, Kinard Lab of Physics, Clemson, SC 29634, USA.
        \and Department of Astronomy, University of Maryland, College Park, MD 20742, USA.
        \and Dipartimento di Fisica, Universita` degli Studi di Roma ‘Tor Vergata’, Via della Ricerca Scientifica 1, I-00133 Roma, Italy.
        \and Istituto Nazionale di Fisica Nucleare, Sezione di Roma ‘Tor Vergata’, Via della Ricerca Scientifica 1, I-00133 Roma, Italy.
        }

   \date{}

% \abstract{}{}{}{}{} 
% 5 {} token are mandatory
 
  \abstract
   {Powerful outflows from active galactic nuclei (AGN) can significantly impact the gas reservoirs of their host galaxies. However, it is still unclear how these outflows can propagate from the very central regions of galaxies to their outskirts, and whether nuclear winds can be driven \textcolor{black}{by} and/or be responsible for drastic spectral transitions.}
   {In this work we test feedback propagation models on the case test of 2MASS 0918+2117 (2M0918), a z=0.149 X-ray variable AGN, which showed tentative evidence for nuclear ultra-fast outflows (UFOs) in a 2005 \textit{XMM-Newton} observation. We also investigate whether UFOs can be related to the observed X-ray variability.}
   {We observed 2M0918 with \textit{XMM-Newton} and \textit{NuSTAR} in 2020 to confirm the presence and characterize the UFOs. We perform a kinematic analysis of the publicly available 2005 SDSS optical spectrum to reveal and measure the properties of galaxy-scale ionized outflows. \textcolor{black}{Furthermore, we construct 20-year-long lightcurves of observed flux, line-of-sight column density, and intrinsic accretion rate from the spectra of the first 4 \textit{SRG}/eROSITA all-sky surveys and archival observations from \textit{Chandra} and \textit{XMM-Newton}}.
}
   {We significantly detect UFOs with v$\sim$0.16c and galaxy-scale ionized outflows with velocities of $\sim$700 km/s. We also find that the drastic X-ray variability (factors >10) can be explained both in terms of variable obscuration and variable intrinsic luminosity.

}
   {\textcolor{black}{Comparing the energetics of the two outflow phases, 2M0918 is consistent with momentum-driven wind propagation. 2M0918 expands the sample of AGN with both UFOs and ionized gas winds from 5 to 6, and brings the sample of AGN hosting multiscale outflows to 19, contributing to a clearer picture of feedback physics. From the variations in accretion rate, column density, and ionization level of the obscuring medium, we propose a scenario that connects obscurers, an accretion enhancement, and the emergence of UFOs.}}


   \maketitle
%
%-------------------------------------------------------------------

\section{Introduction}

Most massive galaxies host at least one supermassive black hole (SMBH) in their inner regions. These SMBHs are known to be the engine of the powerful emitters known as active galactic nuclei (AGN)
%\footnote{In this manuscript we use the acronym AGN both for active galactic nuclei and nucleus.})
, as they accrete gas in their vicinity.
In the past 20 years, observational evidence of tight correlations between the properties of the host-galaxy and the SMBH (e.g \citealp[]{magorrian1998demography,ferrarese2000fundamental, gebhardt2000relationship, haring2004black}), as well as theoretical arguments (e.g \citealp[]{silk1998quasars}) have led the astronomical community to believe that the assembly of the SMBH and the galaxy are connected (see \citealp[]{kormendy2013coevolution} for a review). 

In the AGN/galaxy coevolution framework, the enormous amount of energy that the AGN releases through accretion can severely impact the gas reservoirs of the host
galaxy, possibly quenching or triggering star-formation (e.g \citealp{harrison2017impact}). The so-called AGN "feedback" is now included in most cosmological simulations (e.g. \citealp{di2005energy, sijacki2015illustris, pillepich2021x}), as it is a necessary
ingredient to prevent the formation of galaxies of extreme stellar masses ($\mathrm{M_{\star}} > 10^{12-13} \mathrm{M_{\odot}})$, which are indeed extremely rare in the universe.

A promising way AGN can enact feedback on the host is through
disk winds or outflows that propagate throughout the galaxy (e.g \citealp{fabian2012observational}). 
These winds have been observed on scales that range from
sub-pc to tens of kpc (see e.g. \citealp{cicone2018largely, laha2021ionized}).
They are usually revealed through emission or absorption lines that contain a
blueshifted component, due to the proper motion of the outflowing gas. These features have been observed on galaxy scales in the ionized (optical wavelengths),
molecular (mm/sub-mm wavelengths), or neutral (radio) phases of the ISM, also with
spatially resolved spectroscopy, which reveals complex interactions of the winds with
the surrounding medium (e.g. 
\citealp{brusa2018molecular}, 
\textcolor{black}{\citealp{finlez2018complex}},
\citealp{saito2022kiloparsec}, \citealp{cresci2023bubbles}, \textcolor{black}{\citealp[]{zanchettin2023ngc}}).
The galaxy-scale outflows have typical velocities of $\sim$\textcolor{black}{100-1000} km/s (\citealp{fiore2017agn,musiimenta2023new}).

Higher ionization levels are associated with winds closer to the central launching
engine, the AGN. Broad absorption features in the soft X-rays and absorption lines
above 7 keV ascribed to blended blueshifted Iron lines, respectively known as Warm Absorber
(WA) and Ultra Fast Outflows (UFOs), trace such small-scale outflows. While WAs also have velocities of $\sim$1000 km/s and values of the ionization parameter $\mathrm{\xi}$ in the range log($\mathrm{\xi})\sim 0-2$, UFOs
are characterized by higher ionization states (log($\mathrm{\xi}$) $>$ 4) and extreme, relativistic
velocities (up to 0.3c, \citealp{tombesi2010evidence}).
UFOs are detected in $> 30\%$ of local AGN (e.g. \citealp{tombesi2013unification,igo2020searching, matzeu2023supermassive}), and are expected to generate multiphase, galaxy-scale outflows by interacting with the ISM (\citealp{king2015powerful}). However, whether the energy injected in the \textcolor{black}{outflows} is then efficiently radiated away or instead contributes to their adiabatic expansion is a question still left unanswered by theory. This is the origin of the "momentum-driven" versus "energy-driven" wind propagation mechanism dichotomy (e.g. \citealp{costa2014feedback}).

Observationally, simultaneously measuring outflow energetics on multiple scales for the same source can constrain the mechanism through which winds can propagate from the inner regions to the outskirts of the galaxy. \textcolor{black}{Nevertheless, although the presence of multi-phase large-scale outflows has been observed extensively (e.g., \citealp{herrera2019molecular, shimizu2019multiphase, slater2019outflows, husemann2019close, garcia2021multiphase}}), there are only a handful of sources for which multi-phase and multi-scale outflows have been constrained (see the compilations of \citealp{tozzi2021connecting, bonanomi2023another} and see \citealp{zanchettin2021ibisco}). 
For almost all of the reported sources, by comparing the momentum outflow rates ($\mathrm{\dot{P}}$) of two different wind phases, the mechanism was consistent with one of the two proposed scenarios. Yet, the sample is still too small to derive trends with properties of the AGN.
Moreover, the sample is dominated by sources in which the large-scale outflows are detected in the molecular phase, while only for five sources have the galaxy-scale \textcolor{black}{outflows} been detected in the optical ionized phase. An unbiased sample requires the characterization of large-scale feedback in all phases, and optical observations must catch up to the millimetric studies.

Winds are not only responsible for the large-scale impact of AGN on their host galaxies, but have also been proposed as a possible explanation for the extremely variable changing-look AGN (CL-AGN, see \citealp{ricci2022changing} for a review). In fact, variability is a defining characteristic of AGN, which is observed on timescales that range from minutes to decades (\citealp{padovani2017active}). However, CL-AGN challenge our understanding of what happens at the center of active galaxies.

In the unified model (\citealp{urry1995unified}), the AGN optical spectral classification of Type 1 vs. Type 2 is ascribed to differences in the viewing angle. This is also true in the X-rays, where objects are classified either as obscured (line of sight column density
N$_\mathrm{H} \ > \ 10^{22} \mathrm{cm^{-2}}$) or unobscured (N$_\mathrm{H} \ < \ 10^{22} \mathrm{cm^{-2}}$). Such distinction is due to the presence of a dusty and molecular torus (\citealp{antonucci1985spectropolarimetry, almeida2017nuclear}), which can intercept and block out radiation coming from the very inner regions of the AGN, and give rise to different spectral signatures. 
CL-AGN do not fit into this picture, as they transition on timescales of months to years from one classification to another (e.g. \citealp{miniutti2014properties}, \citealp{yang2018discovery}). 

The existence of CL-AGN requires for the static unified model to include dynamic processes which are not yet fully understood.
 These can take the form of accretion disk instabilities (e.g. \citealp{sniegowska2022modeling}), occultation events (\citealp{risaliti2007occultation}) or ignition/shutdown events (e.g. \citealp{matt2003changing}, \citealp{gezari2017iptf}). In the last few years, an alternative mechanism in which the changing-look event is
explained in terms of variable obscuration due to outflowing gas material has been proposed for some sources (e.g. NGC5548, \citealp{kaastra2014fast}, NGC3227, \citealp{beuchert2015variable} and NGC3783, \citealp{mehdipour2017chasing}).
\textcolor{black}{In addition, since the large-scale outflows are thought to be the result of variable nuclear phenomena, characterizing AGN variability can help us understand the full picture of AGN feedback.} 

In this work, we report a CL-AGN where UFOs play a significant role in driving the X-ray variability. In addition to this, the AGN shows galaxy-scale outflows, which support a momentum-driven scenario.

The paper is organized as follows: In Sect. \ref{sec:2mass} we present the source 2MASS 0918+2117, in Sect. \ref{sec:data} we describe how the data was reduced and in Sect. \ref{sec:opt} we describe the optical spectral analysis used to derive energetics for the ionized outflow. In Sect. \ref{sec:xrayI} we describe the X-ray spectral analysis and in Sect. \ref{sec:xrayII} the X-ray variability. Lastly, in Sect. \ref{sec:disc} we discuss the implications of our results. In this work we assume $\Lambda$CDM Cosmology, with $H_0 = 69.6$, $\Omega_M=0.286$, $\Omega_{\Lambda} = 0.714$.

\section{The case of 2MASS 0918+2117}
\label{sec:2mass}
\begin{table}[t]
\newcolumntype{R}{>{\raggedleft\arraybackslash}X}
\newcolumntype{L}{>{\raggedright\arraybackslash}X}
\newcolumntype{C}{>{\centering\arraybackslash}X}
\def\arraystretch{1.17}
        \begin{tabularx}{0.95\columnwidth}{LL}
            \toprule
            \textbf{Parameter} & \textbf{Value} \\
            \bottomrule
            RA & 09h 18m 48.61s \\
            Dec & +21° 17\arcmin 17.07\arcsec\\
            $z$ & 0.149 \\
            $L_{bol}$ & $2.23\times10^{45} \ \mathrm{erg/s}$\\
            $\mathrm{M_{*}}$ & $7.45\times10^{10} \ \mathrm{M_{\odot}}$\\
            SFR & $0.12 \ \mathrm{M_{\odot}}/\mathrm{yr} $ \\
            \textcolor{black}{$\mathrm{log(M_{BH}/M_{\odot})}$} &$ 7.4 \pm 0.4 $ \\
            \bottomrule
        \end{tabularx}
        \caption{Summary of 2M0918 parameters. The bolometric luminosity $L_{bol}$, the total stellar mass $M_{*}$, and Star Formation Rate (SFR) were all estimated from the SED fitting procedure, described in appendix \ref{app:sed}. \textcolor{black}{The black hole mass $\mathrm{M_{BH}}$ was estimated from the optical analysis presented in Sect. \ref{sec:opt}}}
        \label{tab:2Msum}
\end{table}

\begin{table*}[!ht]
\def\arraystretch{1.17}
\begin{centering}
\newcolumntype{R}{>{\raggedleft\arraybackslash}X}
\newcolumntype{L}{>{\raggedright\arraybackslash}X}
\newcolumntype{C}{>{\centering\arraybackslash}X}
\begin{tabularx}{\textwidth}{LCCCC}
\hline
\textbf{Mission} & \textbf{Obs. ID} & \textbf{Date} & \textbf{Exposure Time (ks)} & \multicolumn{1}{c}{\textbf{Ref.}} \\ \hline
\textit{Chandra} & 2159 & 2001-02-18 & 2 & \cite{wilkes2002x} \\
\textit{XMM-Newton} & 0149170501 & 2003-04-24 & 8 & \cite{wilkes2005xmm} \\
\textit{XMM-Newton} & 0303360101 & 2005-11-15 & 22 & PW07 \\
eROSITA (eRASS1) & - & 2020-05-02 & 0.17 & This Work \\
\textit{XMM} + \textit{NuSTAR}  & 0870820101 & 2020-10-19 & 57+60 & " \\
eROSITA (eRASS2) & - & 2020-11-03 & 0.14 & " \\
eROSITA (eRASS3) & - & 2021-05-06 & 0.14 & " \\
eROSITA (eRASS4) & - & 2021-11-05 & 0.14 & " \\ \hline
\end{tabularx}
\label{tab:lc}
\caption{Observation log of all the X-ray observations used to populate and analyze the X-ray lightcurve of 2M0918. The last column reports the main references that analyzed a given observation previous to this work. Note the low eROSITA exposure times, which are due to its survey observation strategy.}
\end{centering}
\end{table*}

Given the number of open questions and the limited size of the available samples, both regarding CL-AGN and AGN where multiphase outflows are detected, single-object studies can be highly insightful. Detailed multi-epoch spectral analysis can provide constraints on the mechanisms responsible for the observed spectral transitions. Moreover, careful spectral modeling is needed in order to reveal outflows in sources that do not constitute the exceptional bright end of the bulk of the AGN population.

2MASS 0918+2117 (2M0918) is a nearby (z=0.149) Type 1.5 AGN discovered in the Two Micron All Sky Survey (2MASS, \citealp{cutri20022mass}, \citealp{skrutskie2006two}). The most relevant parameters for 2M0918, including values derived from Spectral Energy Distribution (SED) fitting with X-CIGALE (\citealp{yang2022fitting}, see appendix \ref{app:sed}), are presented in Table \ref{tab:2Msum}.

\textcolor{black}{
2M0918 was the object of a total of eight X-ray observations (Table \ref{tab:lc}).
The first three were discussed respectively in \cite{wilkes2002x}, \cite{wilkes2005xmm} and \cite{pounds2007comparison} (PW07 hereafter), and were taken respectively in 2001 with \textit{Chandra}, in 2003 and 2005 with \textit{XMM-Newton}.
We also followed up 2M0918 with \textit{XMM-Newton} + \textit{NuSTAR} in 2020 for a total of $\sim 57+60 \ $ks.
Additionally, 2M0918 was also observed by eROSITA (\citealp{Predehl_2021}, P21), aboard the Spectrum-Roentgen-Gamma (\textit{SRG}) observatory (\citealp{sunyaev2021srg}) once every six months in the first 4 All-Sky Surveys (eRASS1-4, \citealp{Merloni_2021}, \textcolor{black}{\citealp[]{merloni2024srg}}).
}
\textcolor{black}{The 2001 \textit{Chandra} observation did not show significant obscuration, while the spectrum obtained in the 2003 observation appeared both fainter (factor $\sim 5$) and harder, suggesting an increase in the column density along the line of sight. The 2005 spectrum appeared 10 times brighter and much softer than in 2003. PW07 explained the variability in terms of changes in the intrinsic luminosity, tracing the accretion disk/corona. However, the changes in the hardness of the spectra suggest that variable obscuration could also be at play. PW07 also noted the presence of an absorption feature above
7 keV, interpreted in the work as tracing an Ultra Fast Outflow of $\mathrm{V_{OF}}\sim0.15c$, despite the tentative evidence and the term "UFO" not being common in literature yet. 
Blue-winged oxygen and Balmer emission lines in the 2005 SDSS optical spectrum confirm that the AGN is caught in its active feedback phase, as shown in Sect \ref{sec:opt}.
}

\textcolor{black}{
In this work, we shed light on the 
nature of the X-ray variability of 2M0918, as well as characterize the multiphase and multiscale winds of which previous studies show evidence. 
In order to characterize the UFO we make use of our \textit{XMM-Newton} and \textit{NuSTAR} follow-up, and concerning the ionized outflows, we analyze the publicly available SDSS optical spectrum. All of the X-ray data were used to describe the variability.
}
\section{Data reduction}
\label{sec:data}

\begin{figure*}[t]
    \centering
    \includegraphics[width=\textwidth]{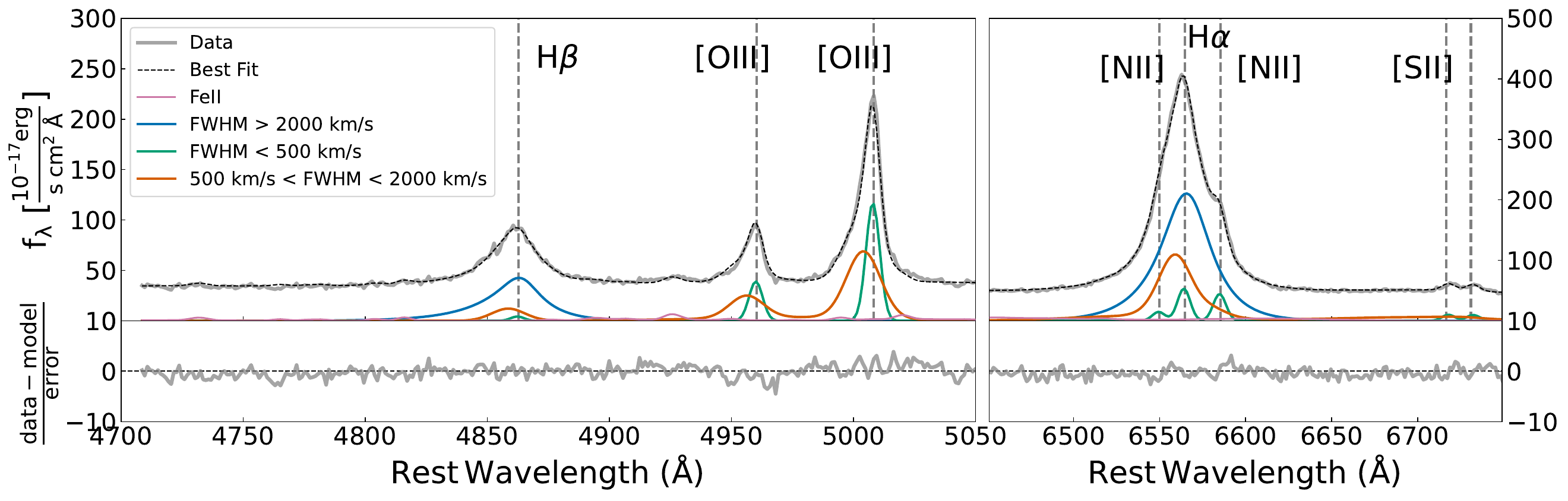}
    \caption{Best fit model of the H$\beta$ and  H$\alpha$ complexes. The original dataset is grey, the total line emission is the dashed black line, and the pink curve is the FeII contribution. Different color lines represent the multiple emission line components described in the text. The orange Gaussian lines clearly reveal the presence of a blue-shifted component.
    }
    \label{hbeta}
\end{figure*}

SDSS observed 2M0918 in November 2005. For our optical analysis, we downloaded the spectrum from the publicly accessible catalog \footnote{https://dr12.sdss.org/}.

The X-ray observations, listed in Tab. \ref{tab:lc}, were reduced differently depending on the instrument. The 2003, 2005 and 2020 \textit{XMM-Newton} observations have the longest exposure times of all the available X-ray observations. For all three observations, the pn (\citealp{struder2001european}), MOS1 and MOS2 (\citealp{turner2001european}) data were processed using the standard \textsc{SAS v.20.0.0} procedures. We then extracted the spectra following the procedures of \cite{piconcelli2004evidence} and \cite{bianchi2009caixa}, which maximize the S/N in a given band. We chose the hard 2-10 keV band, as this is where UFO features should be found, and we only selected single and double pixel events (pattern 0–4 and 0–12 for pn and MOS, respectively). The S/N optimization works by testing combinations of multiple source extraction radii and background thresholds, given a fixed background extraction region. Using a background region of 50", we found that the S/N is maximized in the 2-10 keV band for pn, MOS1, and MOS2 extraction regions with radii of 33", 32” and 38”  respectively in 2003, 40”, 30” and 40” in 2005, and 17”, 20” and 15” in 2020. These correspond to encircled energy fractions (EEF) of $\sim$80-90$\%$ in 2003 and 2005 and $\sim$60$\%$ in 2020. 
The response files were then generated with the
\textsc{sas} tasks \textsc{rmfgen} and \textsc{arfgen} with the calibration EPIC files version \textsc{v3.13}.

The 2020 \textit{NuSTAR} observations were taken 3 days after the \textit{XMM-Newton} pointing. The data of the two cameras (FPMA and FPMB) were processed with the \textit{NuSTAR} Data Analysis Software package (NuSTARDAS) version 2.1.2 within HEAsoft v.6.30. We used the calibration files in the \textit{NuSTAR} CALDB (version 20220510) and produced clean event files using the task \textsc{Nupipeline} with standard filtering criteria. We did not detect significant solar flares through the \textsc{nustar\_filter\_lightcurve} IDL script.
We selected an extraction region with a radius of 40" ($\sim$ EEF), and simulated the background spectrum at source location using the \textsc{NuSkyBgd} IDL script (\citealp{wik2014nustar}), which accounts for the position-dependent stray light component.

We binned the \textit{XMM-Newton} spectra using the optimal binning option (KB, \citealp{kaastra2016optimal}) in the HEAsoft \textsc{ftgrouppha} task. For all three observations, the pn, MOS1 and MOS2 spectra were analyzed simultaneously, with the addition of the FPMA and FPMB spectra in the 2020 spectral set. We binned the \textit{NuSTAR} spectra to one count per bin to preserve spectral resolution.

For the 2001 \textit{Chandra} data, we used products as extracted by the CXC’s Automated
Processing pipeline. The data were processed with CXC software version 10.9.2, using CalDB version 4.9.4. We then extracted the source and background spectra, and produced the ARF and RMF with the \textsc{specextract} task of the \textit{Chandra} Interactive Analysis of Observations (CIAO) software version 4.14.0, using circular regions with radii of 5” ($>90\%$ EEF) for the source and 27” for the background. We binned the spectrum to one count per bin.
We adopt Cash statistics (\citealp{cash1979parameter}) for all of the abovementioned X-ray observations. \textcolor{black}{Our goodness-of-fit estimator is the maximum-likelihood-based C-stat value, which is approximately distributed like the ${\chi}^2$ estimator used in Gaussian statistics. Therefore, a good fit will be quantified by a ratio of C-stat over the degrees of freedom close to 1, similarly to the reduced ${\chi}^2$. }

To analyze the eROSITA spectra we first matched 2M0918 with the eROSITA catalogs for eRASS1 through eRASS4 (available within the eROSITA-DE consortium) with \textsc{topcat} v4.8-6. For all 4 catalogs, the matching was positive and unambiguous.
We subsequently downloaded the source products from the web tool \textsc{DATool} as extracted by the eROSITA Science Analysis Software System (eSASS, \citealp{brunner2022erosita}) pipeline in the latest available configuration (v. 211214). The products included lightcurves, ARFs, RMFs, and source and background spectra for each eRASS and for different Telescope Module (TM) combinations (see \citealp{liu2022erosita} for details on how the source and background extraction regions are defined in the automated pipeline).
We simultaneously analyzed the spectra of the combined TM1, TM2, TM3, TM4, and TM6, leaving out TM5 and TM7 because these modules are known to be affected by light leaks, which contaminate observations (see P21). Due to the survey nature of the eROSITA observational strategy, exposure times are of the order of $\sim 200$s, which correspond to photon counts $<40$ for 2M0918. For this reason, we do not bin the spectra and apply Bayesian methods, suited for low-photon statistics, which will be presented in Section \ref{sec:xrayII}.

\section{Optical spectral analysis}
\label{sec:opt}

We modeled the spectrum in two steps. First we fitted the continuum with a power-law and the FeII empirical template presented in \cite{veron2004unusual}. The initial fit covered the entire wavelength range but masked the prominent broad-line-region (BLR) and narrow-line-region (NLR) lines. 
We then proceed to simultaneously fit the emission lines and the continuum using the template derived from the previous fit. Regarding the lines, we fitted eight narrow (FWHM $<$ 500 km/s) Gaussian lines to account \textcolor{black}{for the NLR emission of} H$\alpha$, H$\beta$ and the [OIII]$\lambda4958,5007$, [NII]$\lambda6548,6584$ and [SII]$\lambda6716,6730$ doublets. All widths and velocity shifts were tied together, and the intensity ratios between the components of the [OIII] and [NII] doublets were set to 1:2.99, assuming typical gas conditions (\citealp{osterbrock1981seyfert}). We also model the broad (FWHM $>$ 2000 km/s) H$\alpha$ and H$\beta$ lines originating from the BLR with broken powerlaws convolved with Gaussian functions (e.g. \citealp{cresci2015blowin, perna2021physics}). The broken power law indices were tied between the two lines. 
Additionally, we included six moderately broad (500 km/s $<$ FWHM $<$ 2000 km/s) Gaussian components (H$\alpha$, H$\beta$, [OIII] and [SII] doublets) to account for turbulent kinematics or outflows. The widths, velocity shifts, and relative intensities were constrained as above. . \textcolor{black}{All fitted emission lines have a signal-to-noise ratio of 8 or above}.

Figure \ref{hbeta} shows the result of our fitting procedure in the H$\alpha$ and H$\beta$ regions. As shown by the orange curves in the plot, the outflowing component contributes significantly to line emission and cannot be excluded by our model

\subsection{Ionized gas kinematics and black hole mass}

The prescriptions we used for characterizing the outflow kinematics have been established in the literature for over a decade (e.g. \citealp{cano2012observational}, \citealp{fiore2017agn}).
By assuming spherical geometry of the outflow, the mass outflow rate can be computed as $\mathrm{\dot{M}_{OF}} = 3 \times \mathrm{V_{OUT}} \times \mathrm{M_{OF}} \times \mathrm{R_{OF}^{-1}}$, where $\mathrm{V_{OUT}}$ is the maximum outflow velocity, $\mathrm{R_{OF}}$ is the outflow radius and $\mathrm{M_{OF}}$ is the mass of the gas entrained in the outflow.

Without spatially resolved spectroscopy $\mathrm{R_{OF}}$ can only be constrained as an upper limit. Knowing that the fibers on the SDSS spectrograph are 3" in diameter, the observed outflow is contained within a $3.9 \ \mathrm{kpc}$ radius at the redshift of this source.

\textcolor{black}{Following, \cite{cano2012observational},} the mass $\mathrm{M_{OF}}$ can be estimated from the integrated luminosity of the \textcolor{black}{broad} [OIII]$_{\mathrm{\uplambda5007}}$ line:
\begin{equation}
    \mathrm{M_{OF}} = 5.3 \times 10^7 \frac{L_{44}(\mathrm{[OIII])}}{n_{e_3}10^{[\mathrm{O/H}]}} \ \mathrm{M_{\odot}},
\end{equation}
where $L_{44}(\mathrm{[OIII]})$ is the luminosity of the \textcolor{black}{outflow component} of the [OIII]${\lambda 5007}$ line in units of $10^{44}$ erg/s, $10^{[\mathrm{O/H}]}$ is the metallicity of the outflowing gas and $n_{e_3}$ is the electron density of the same gas in units of 1000 cm$^{-3}$.

We estimated $n_{e_3} = 0.16 $ from the ratio R=1.24 of the two outflowing components of the [SII] doublet (assuming a gas temperature of $10^4$ K, \citealp{osterbrock1981seyfert}), and we assumed solar metallicity. We measured the observed $L_{44}(\mathrm{[OIII]}) = (7.9\pm0.2)\times10^{-3}$.
However, this value is affected by host-galaxy extinction, which we quantified through the Balmer decrement method (\citealp{osterbrock2006astrophysics}).
The observed Balmer ratio of the outflowing component is  H$\alpha$/H$\beta = 3.9^{+0.2}_{-0.4}$. We adopted the \cite{calzetti2000dust} extinction curve and, for an intrinsic H$\alpha$/H$\beta$ ratio of 2.86, we measured E(B-V)$=0.30^{+0.04}_{-0.10}$ mag. 
The extinction-corrected [OIII] luminosity is therefore $L_{44}(\mathrm{[OIII]}) = (2.1^{+0.3}_{-0.6}) \times 10^{-2}$. 

For $\mathrm{V_{OUT}}$, we used the common non-parametric maximum velocity estimator $v_{10}$, which corresponds to the 10th percentile of the overall line profile. We obtained $\mathrm{V_{OUT}}\sim700$ km/s for [OIII]$_{\mathrm{\uplambda 5007}}$. The derived mass outflow rate is then $\mathrm{\dot{M}_{OF}} = 3.7^{+0.5}_{-1.1} \ \mathrm{M_{\odot}}$/yr, and the momentum 
 outflow rate $\mathrm{\dot{P}}$ and kinetic power $\mathrm{\dot{K}}$ are simply $\mathrm{\dot{P}} = \mathrm{\dot{M}_{OF}}\times \mathrm{V_{OUT}} =  (2.1^{+0.3}_{-0.6}) \times 10^{-1} \ L_{bol}/c$ and $\mathrm{\dot{K}} = \frac{1}{2} \times \mathrm{\dot{P}\times V_{OUT}} = (5.6^{+0.8}_{-0.2}) \times 10^{41}$ erg/s.

 We also performed a similar computation from the H$\alpha$ line (see e.g. \citealp{cresci2023bubbles}). Using hydrogen lines has the advantage of releasing the assumption on the metallicity of the gas. However, the derived parameters are affected by the higher degree of degeneracy between the many emission components (i.e. the presence of BLR emission). With an observed \textcolor{black}{broad} H$\alpha$ luminosity of $L_{44}(\mathrm{[H\alpha]}) = (5.6^{+0.4}_{-0.1}) \times 10^{-3}$ we obtained $\mathrm{\dot{M}_{OF}} = 17.4^{+0.9}_{-1.9} \ \mathrm{M_{\odot}}$/yr, $\mathrm{\dot{P}} = (1.0 \pm 0.1) \,L_{bol}/c$ and $\mathrm{\dot{K}} = (2.6^{+0.1}_{-0.3}) \times 10^{42}$ erg/s. The energetics derived from H$\alpha$ are larger by a factor 4.7 than the values obtained from [OIII], \textcolor{black}{in line with the results by \cite{fiore2017agn}.}

We measured the mass of the SMBH through the single-epoch virial method, which is based on the empirical $R_{BLR} \propto L^{0.5}$ relation discovered through reverberation mapping studies (e.g. \citealp{kaspi2005relationship}). We used the relation presented in \cite{dalla2020sloan} (Eq. 38), which makes use of the broad H$\beta$. 
The Balmer ratio in the BLR is H$\alpha$/H$\beta = 8.58^{+0.45}_{-0.32}$, so the intrinsic luminosity is $L_{44}(\mathrm{[H\beta]}) = (1.7^{+0.2}_{-0.3}) \times 10^{-1}$. With a line dispersion of $\sigma\sim1100$ km/s, this corresponds to a mass of $\mathrm{log(M_{BH}/M_{\odot}) = 7.4 \pm 0.4 }$, where the errors are conservatively estimated to be 0.4 dex, based on the argument presented in \cite{shen2013mass}.
%
%\begin{equation}
%    log\bigg(\frac{M_{BH}}{M_{\odot}}\bigg) = A + B \ log\bigg(\frac{\lambda \ L_{\lambda}}{10^{42} \ \mathrm{erg /s}}\bigg) + 2 log \bigg(\frac{FWHM}{1000 \ \mathrm{km/s}}\bigg)
%\end{equation}
%Where $A$ and $B$ are constants and $\lambda L_{\lambda}$ is the unextinct luminosity calculated at $\lambda$. 
%Following the prescriptions of \cite{greene2005estimating}, we use the FWHM of H$\beta$ and estimate the continuum luminosity at 5100$\AA$ (A=6.64, B=0.64).
%This corresponds to a mass of $log(M_{BH}/M_{\odot}) = 7.6 \pm 0.4 $. 
%The broad H$\beta$ luminosity can also be used as a proxy of the continuum luminosity, as it correlates with $L_{5100\AA}$. When we implement this method (A=6.56, B=0.56) we measure a black hole mass of $log(M_{BH}/M_{\odot}) = 7.4 \pm 0.4 $. On both values, the errors are conservatively estimated to be 0.4 dex, based on the argument presented in \cite{shen2013mass}.
%As both values are close to each other, for the rest of this work, we use the average $log(M_{BH}/M_{\odot}) = 7.5 \pm 0.5 $. 

\section{X-ray spectral analysis I: searching for winds}
\label{sec:xrayI}

\begin{figure}[!t]
    \centering
    \includegraphics[width=\columnwidth]{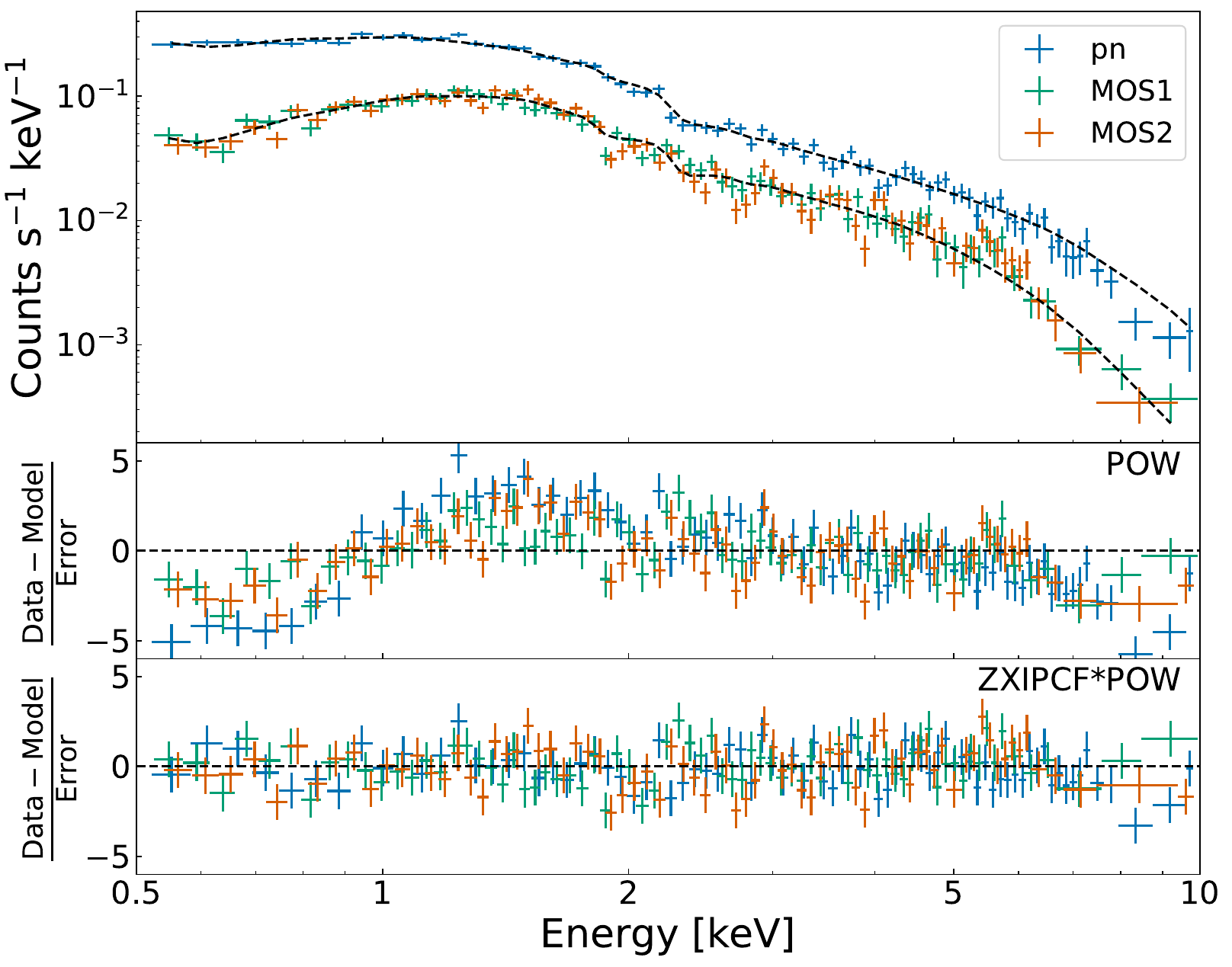}
    \caption{2005 EPIC spectra of 2M0918. The top panel shows the three spectra, the middle panel shows residuals for the power-law-only model, while the bottom panel shows residuals for the selected absorbed-power-law model.}
    \label{onlypow2005}
\end{figure}

We reanalyzed the 2005 \textit{XMM-Newton} spectrum (shown in Fig. \ref{onlypow2005}) first presented in PW07, in order to confirm the UFO feature. We do so through the \textsc{Xspec} spectral fitting package v.12.12.0 (\citealp{arnaud1996xspec}) by first identifying the best continuum model, and subsequently looking for excess emission or absorption features in the hard band.

We first fit the three EPIC spectra simultaneously with a simple Galactic absorption and power-law model (\textsc{Xspec: const\footnote{When fitting multiple spectra the addition of a constant is needed in order to account for different instrumental normalization.}*tbabs*powerlaw}) between 0.5 and 10 keV, in order to model the continuum. We use a galactic absorption of $\mathrm{N_{H}}=4 \times 10^{20}$ cm$^{-2}$, as derived from the \textsc{HEAsoft} task \textsc{NH} (\citealp{kalberla2005leiden}).
This initial model is insufficient at reproducing the whole spectrum, with a C-stat value of 817 over 273 degrees of freedom (d.o.f.). The residuals (Fig. \ref{onlypow2005}, middle panel), show a deficit of counts below 1 keV and a nexcess in the 1-2 keV band, suggesting the presence of an absorber in the data.

We included in our model a photoionized and partially covering absorber to the model (\textsc{zxipcf}) to account for the deviations in the soft band. The residuals are less scattered (Fig \ref{onlypow2005}, bottom panel), and the fit highly improves (C-stat: 329 over 270 d.o.f.). The fit parameters are shown in Tab. \ref{tab:xmmfit}.
The value of the absorber's $\mathrm{N_{H}}$ is very mild, while the ionization parameter is consistent with a neutral obscuring medium. 
This is our base continuum model for the rest of the following analysis.

We then proceeded to look for hard absorption and emission features, through a \textit{line scan}, as first proposed in \cite{cappi2009x}. The procedure, which has been extensively used in literature (e.g. \citealp{tombesi2010evidence, bertola2020x, matzeu2023supermassive}) operates as follows: The baseline continuum model is fit and its C-stat value is stored. A narrow, unresolved line (\textsc{Xspec: zgauss}, with $\sigma=10$ eV) is added to the model, free to vary in both positive and negative normalization, to account for both emission and absorption features. The line energy is shifted 100 times between 5 and 10 keV (rest-frame), and, for each of these energies, normalization is allowed to vary both positively and negatively 100 times as well in a range visually selected. The C-stat value of each of these 100x100 combinations is stored. Contour plots are then produced for values of $\Delta \mathrm{C}$ of -2.3, -4.61, -5.99,  -9.21, which correspond to 68\%, 90\%, 95\%, and 99\% confidence level fit improvement. 

%The resulting plot (which is shown in Fig. \ref{fig:cappi2005}) is to be read inversely compared to "standard" contour plots: inner closed lines indicate a higher significance of fit improvement. 

\begin{figure}[t]
    \centering
    \includegraphics[width=\columnwidth]{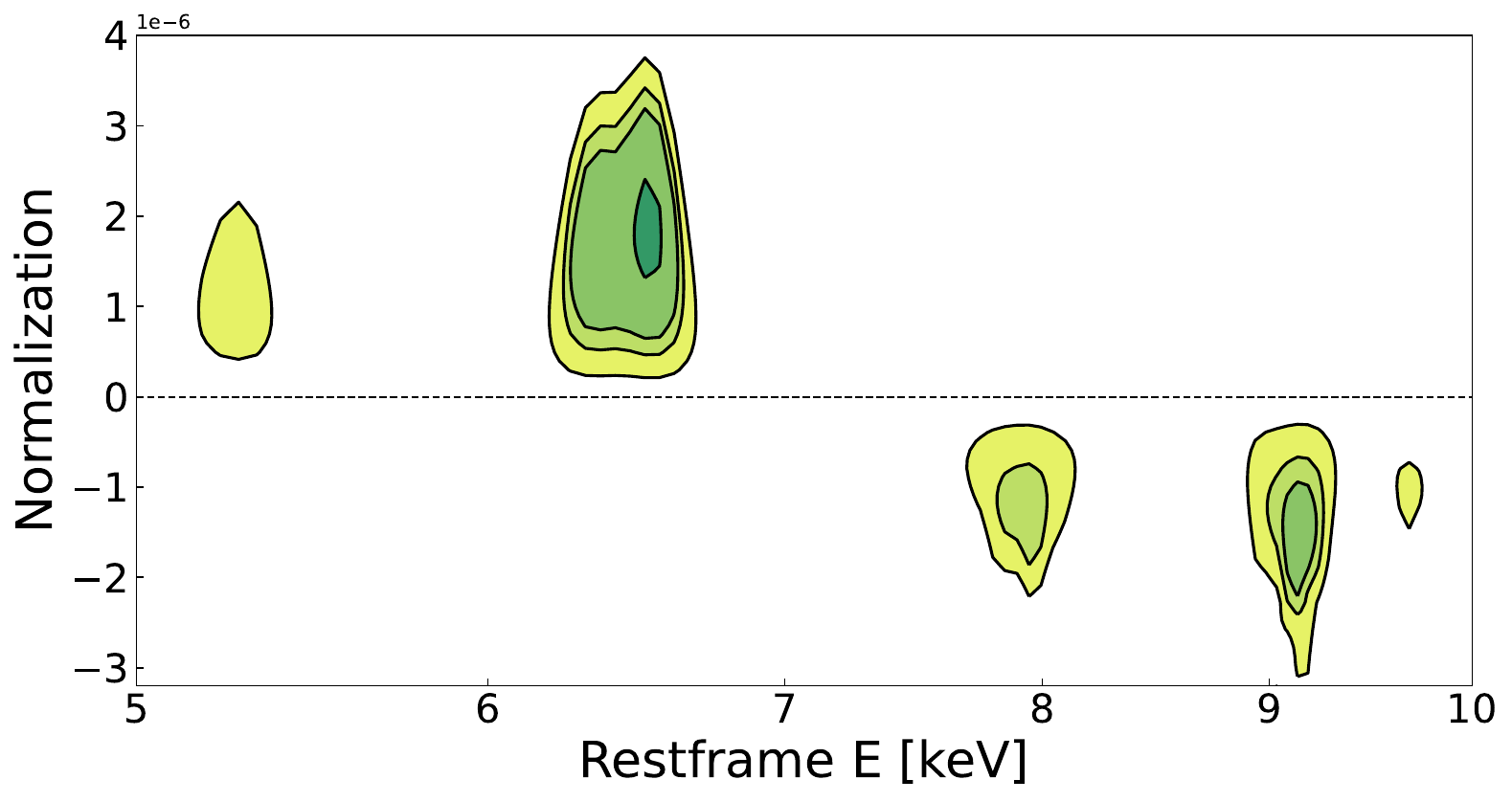}
    \caption{Line scan contours. The vertical axis corresponds to the line normalization values and the horizontal to the line energy in the source rest frame. the contours correspond to 68\%, 90\%, 95\% and 99\%
confidence level fit improvement, as the color gets darker.}
    \label{fig:cappi2005}
\end{figure}

Guided by the results shown in Fig. \ref{fig:cappi2005}, we then fitted a narrow emission line between 6 and 7 keV. The fit significantly improved (99\%)  and the line was found at E=$6.5^{+0.1}_{-0.2}$ keV, with an equivalent width of $0.10\pm0.04$ keV. The inclusion of this line in the model improves the C-stat to 319 over 268 d.o.f., which corresponds to an F-test significance of 98\%. As this line is compatible with the known Iron K${\alpha}$ transition (E=6.4 keV), we include it in our baseline model. 

% \usepackage{booktabs}
\begin{table*}[t]
\def\arraystretch{1.17}
\begin{center}
\newcolumntype{R}{>{\raggedleft\arraybackslash}X}
\newcolumntype{L}{>{\raggedright\arraybackslash}X}
\begin{tabularx}{1\textwidth}{ccLLc}
\toprule
\textbf{Observation}              & \textbf{Component}                 & \textbf{Parameter}                 & \textbf{Value}                       & \textbf{Units}      \\ \midrule
\textbf{2005} & \textsc{zxipcf}   & N$_{\mathrm{H}}$                   & $(5.2\pm0.1) \times 10^{21}$         & cm$^{-2}$           \\
                                  &                                    & log($\xi$)                         & $<-0.19$                            & -                   \\
                                  &                                    & Covering fraction                  & $0.84\pm0.03$                        & -                   \\
\textbf{}                         & \textsc{powerlaw} & Photon index                       & $2.07\pm0.07$                        & \textbf{-}          \\
\textbf{}                         & \textbf{}                          & Normalization                      & $(5.8\pm0.6) \times 10^{-4}$         & photon/keV/cm$^{2}$/s \\ \midrule
\textbf{2020} & \textsc{zxipcf}   & N$_{\mathrm{H}}$                   & $5.0^{+0.6}_{-0.7} \times 10^{22}$ & cm$^{-2}$           \\
                                  &                                    & log($\xi$)                         & $1.1^{+0.2}_{-0.4}$                  & -                   \\
                                  &                                    & Covering fraction                  & $0.81^{+0.05}_{-0.07}$               & -                   \\
                                  & \textsc{powerlaw} & Photon index & $2.11^{+0.18}_{-0.15}$               & -                   \\
                                  &                                    & Normalization                      & $(1.7\pm0.5) \times 10^{-4}$         & photon/keV/cm$^{2}$/s \\ \bottomrule
\end{tabularx}
\end{center}
\caption{Values for best-fit continuum models for the 2005 and 2020 \textit{XMM-Newton} and \textit{NuSTAR} observations. Significant variations are present in the column density, ionization parameter and the power-law normalization (intrinsic luminosity). All errors are at 1$\upsigma$ level.}
\label{tab:xmmfit}
\end{table*}

\begin{figure}[t]
    \centering
    \includegraphics[width=\columnwidth]{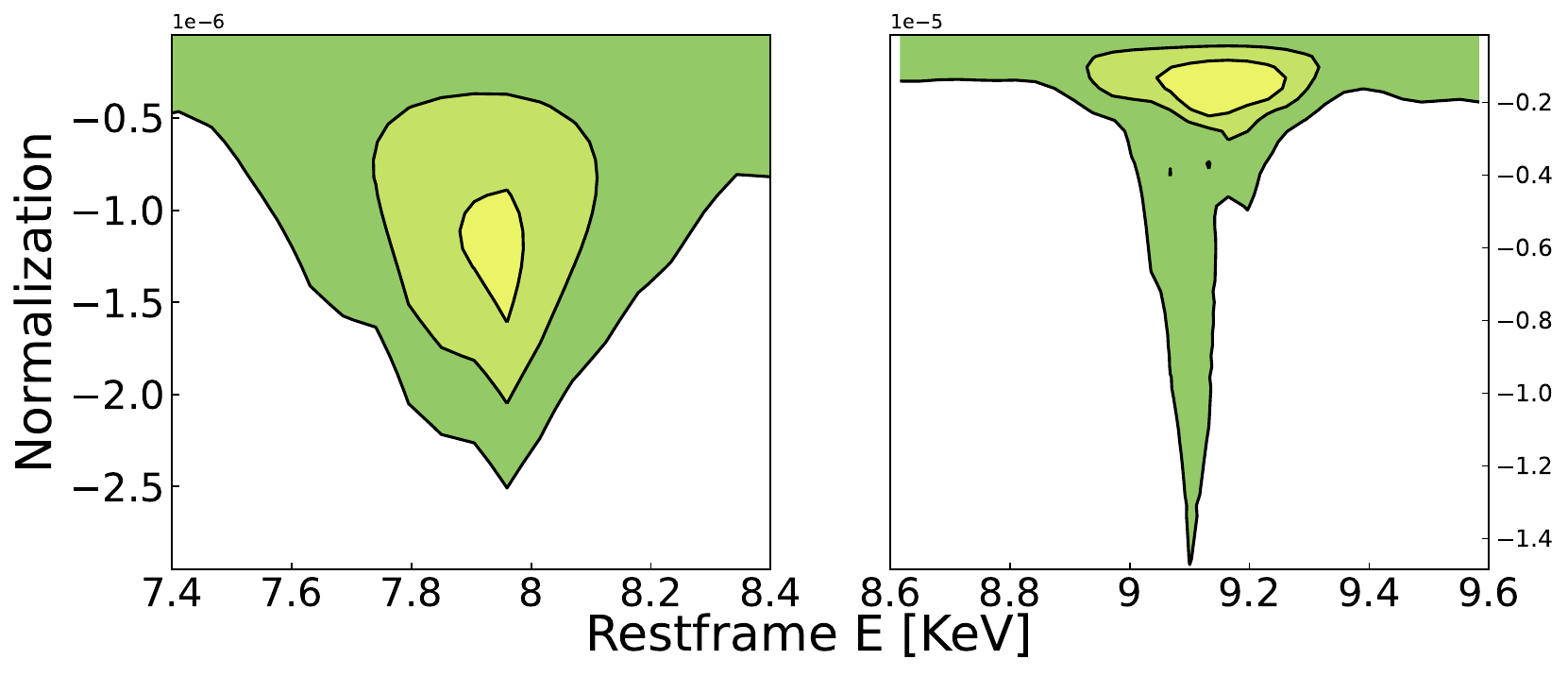}
    \caption{Contours plots of the absorption features of the 2005 2M0918 \textit{XMM-Newton} spectra. The lines correspond to 68\%, 90\% and 99\% significance levels. Note that the 99\% line does not close in either case.}
    \label{fig:abscont2005}
\end{figure}

The two hard absorption features are less significant. Fitting one narrow Gaussian component, we find a line at E=$9.1 \pm 0.1$ keV, with a fit improvement of $\Delta\mathrm{C/d.o.f.} = 7/2$, which corresponds to an F-test significance of 95 \%. If we move the initial value of the line energy towards the second $\sim$ 8 keV feature, the new local minimum is at E=8.0$\pm 0.2$ keV. The improvement is even less significant ($\Delta\mathrm{C/d.o.f.} = 5/2 \rightarrow $ F-test: 87\%). The contours for both of these lines are shown in Fig. \ref{fig:abscont2005}, where it is evident that the 99\% curve does not close, as is consistent with null normalization.

\begin{figure}[t]
    \centering
    \includegraphics[width=\columnwidth]{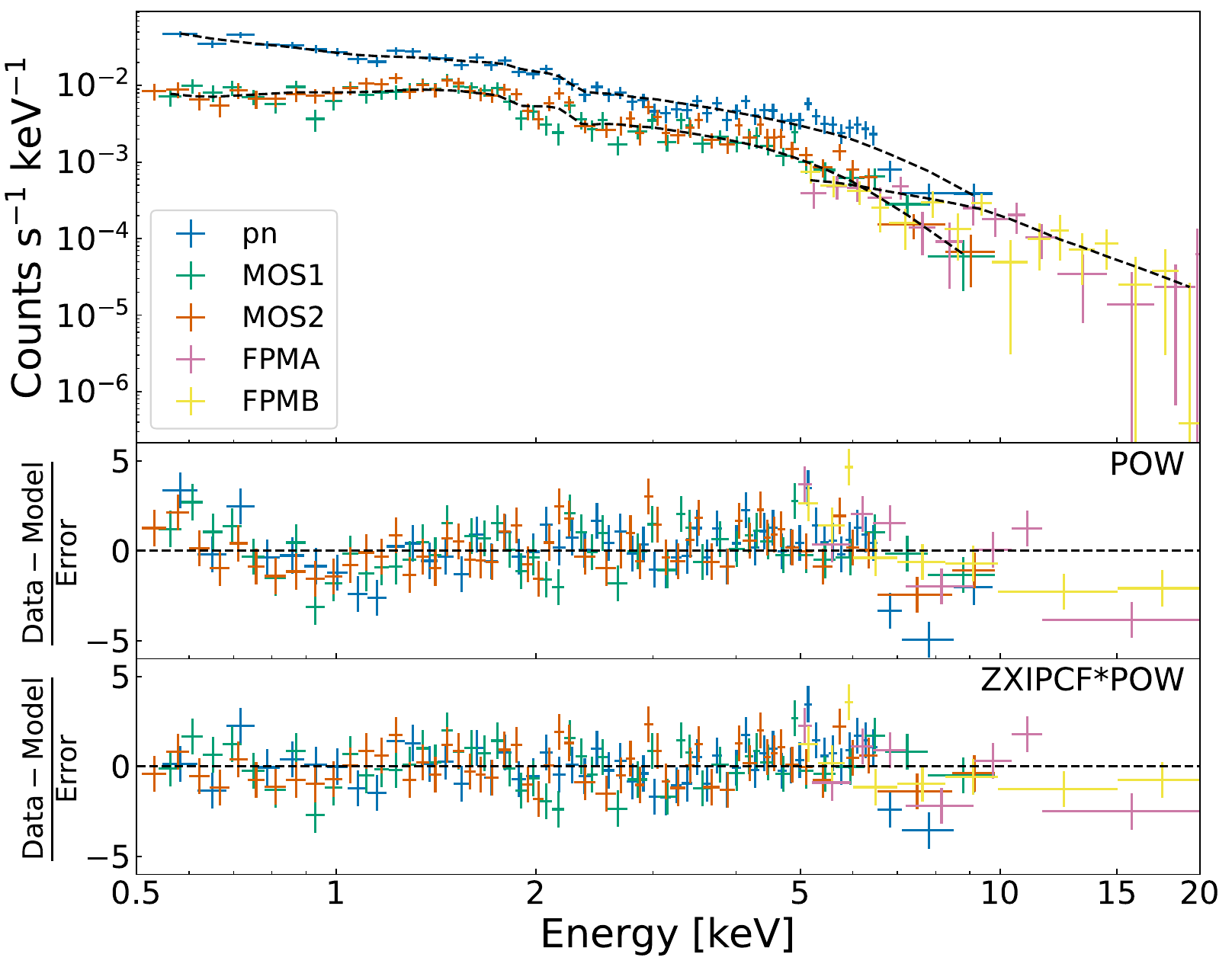}
    \caption{2020 \textit{XMM-Newton} and \textit{NuSTAR} spectra of 2M0918. The first panel shows the five spectra, the second panel shows residuals for the power-law-only model, and the last panel shows residuals for the best model. }
    \label{onlypow2020}
\end{figure}

\medskip
We perform a similar analysis on the 2020 \textit{XMM-Newton} and \textit{NuSTAR} spectra, which are shown in Fig. \ref{onlypow2020}. Following the same approach as in the 2005 case, but extending the energy range up to 20 keV, we find once again that the source is best reproduced by a \textsc{const*tbabs*zxipcf*powerlaw} model. In fact, the simple power-law model has a value of C-stat/d.o.f. of 560/521, which improves to 490/518 with the inclusion of the absorber.

Despite the model being the same, some parameters show significant changes from
the 2005 to the 2020 observations: the absorber, which was previously consistent with
not being ionized, has increased its ionization parameter $\xi$ of more than one order of
magnitude, and the column density also increased by a factor $\sim10$. The intrinsic luminosity measured in the 2-10 keV band through the \textsc{clumin} task also decreased
significantly: log(L$_{2-10,\ 2020}$) = 43.37 $\pm$ 0.01 vs log(L$_{2-10,\ 2005}$) = 43.91 $\pm$ 0.01.
Covering fraction and photon index stayed consistent between the two observations.

\begin{figure}[]
    \centering
    \includegraphics[width=\columnwidth]{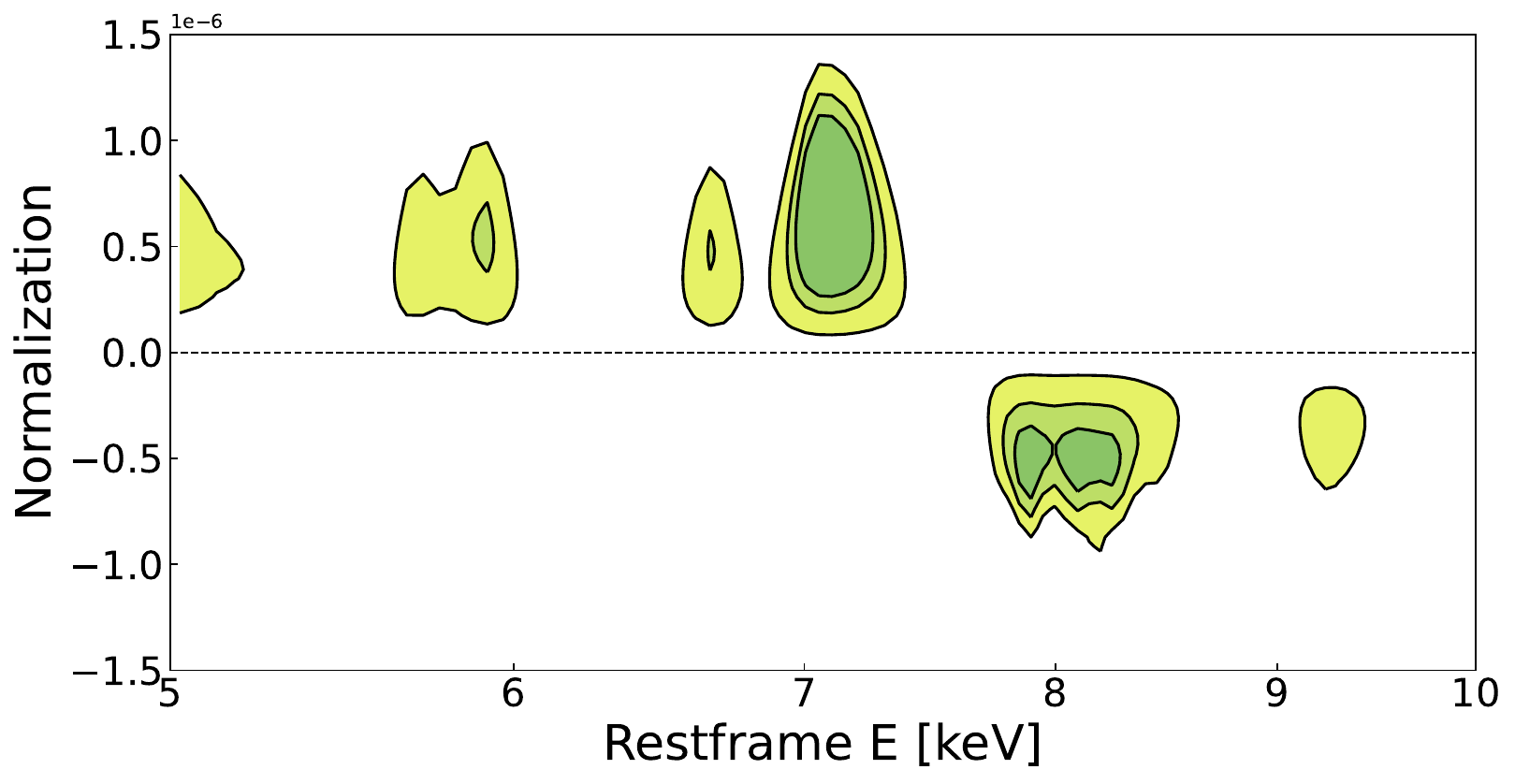}
    \includegraphics[width=\columnwidth]{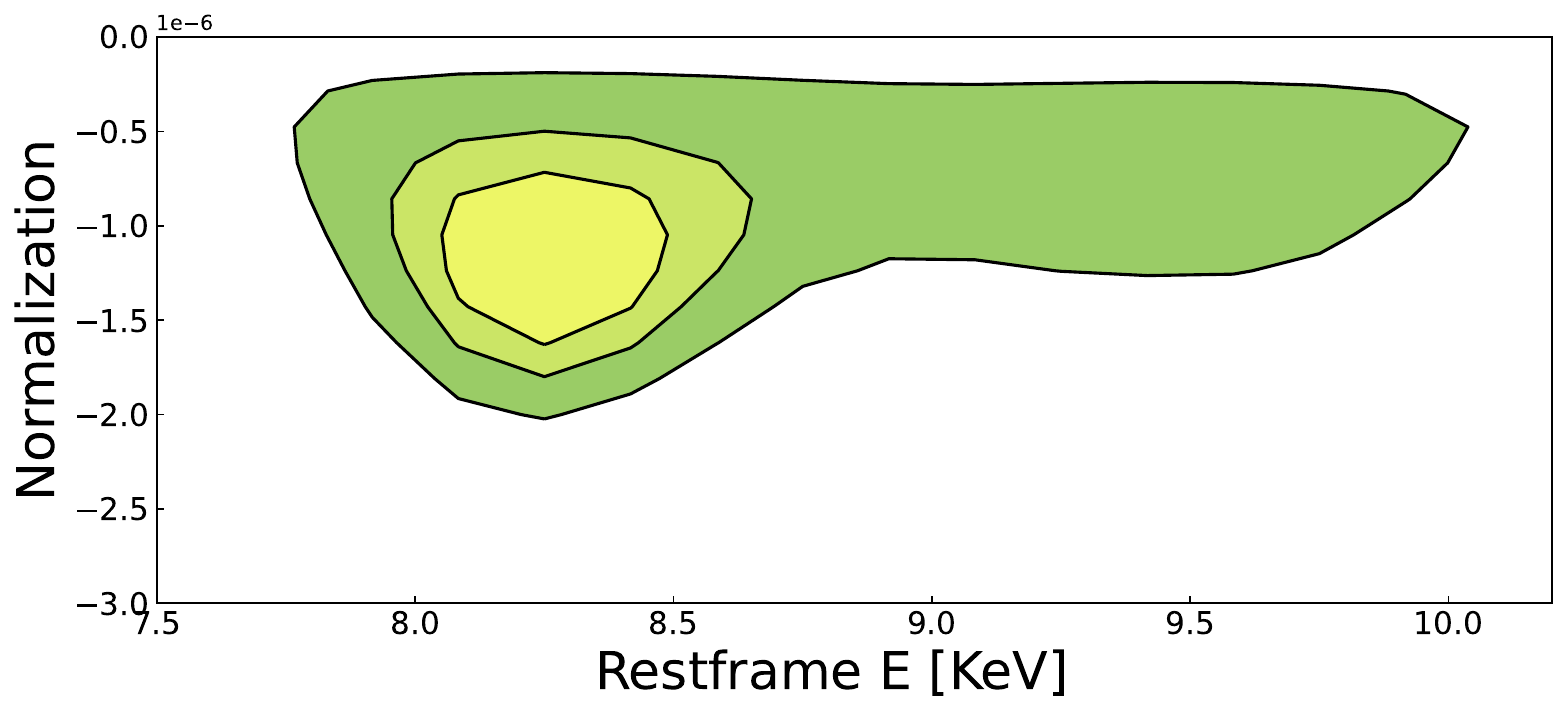}
    \caption{Top: line scan contours of the 2020 observations. Bottom: contours plots of the absorption feature of the 2020 \textit{XMM-Newton} spectra modeled as a broad line. Both figures are to be respectively interpreted as in Fig. \ref{fig:cappi2005} and Fig. \ref{fig:abscont2005}.}
    \label{fig:cappi2020}
\end{figure}

We repeat the same line scan procedure previously performed on the 2005 spectra, and obtain the contours shown in Fig. \ref{fig:cappi2020}, top panel for the 2020 spectra. The iron k$\alpha$ line at 6.4 keV is no longer present, although a less significant emission line at 7.1 keV is evident. At around 8 keV, the contours reveal the presence of a broad absorption feature which appears to be due to the blending of 2 different lines. However, when the two features are modeled independently, neither is very significant ($\Delta\mathrm{C}/d.o.f = 6/2$) and revealing their separation is challenging. A more robust fit is obtained when the feature is modeled as one single broad Gaussian. By allowing $\upsigma$ to vary, we find a line at E=$8.2^{+0.6}_{-0.3}$ keV and $\upsigma = 0.47^{+0.8}_{-0.2}$ keV, with a statistic improvement of $\Delta\mathrm{C}/d.o.f = 11/3 \rightarrow $ F-test: 99.2\%, Fig. \ref{fig:cappi2020}, bottom panel. 

\subsection{Wind significance assessment}
\label{sec:windsig}
As shown in \cite{protassov2002statistics}, the F-test should not be taken at face value when estimating the significance of emission/absorption lines, as it can lead to overestimations. In order to robustly quantify whether the features are physical or are due to random fluctuations of the spectra, we made use of Monte Carlo simulations as follows. 

%\begin{figure*}[!t]
%    \centering
    %\includegraphics[width=0.95\textwidth]{Figures/sighist.pdf}
    %\caption{Results of the Monte Carlo simulations to assess the significance of each potential UFO feature. The x-axis represents the C-stat difference 
    %($\Delta C$) for the simulated data, as described in the text. The vertical dashed line represents the $\Delta C$ on the observed data. As one can see, only the 2020 broad feature is significant. It can also be noted that the mode of the distribution shifts to lower values for Monte Carlo simulations of broad features, as the alignment of photon fluctuations needed to emulate a line gets more exceptional.}
    %\label{fig:sighist}
%\end{figure*}

We used the continuum models in Tab. \ref{tab:xmmfit}, with parameters set to the ones found in our analysis, to simulate $3\times5000$ spectum and background pairs for the 2005 observations (pn, MOS1, and MOS2), and $5\times5000$ pairs for the 2020 data (pn, MOS1, MOS2, FPMA, and FPMB). We also included the 6.5 keV emission line in the 2005 continuum.
After binning the simulated data with the same criteria as in the analysis, we fitted the 2 sets of spectra with the same models used to produce them. We stored the C-stat value and the number of d.o.f. of each fit.
We then re-fitted the spectra with an added Gaussian line, free to vary between 4 and 10 keV (source rest-frame) and to be in either absorption or emission. The line was stepped through all of its allowed energy range, in order to find the absolute fit minimum. We did so as our interest is in how often random fluctuations are mistaken for spectral lines \textit{in general}, and not just limited to our specific energy and normalization. We stored once again the C-stat value and d.o.f.
We then counted the instances where the inclusion of the Gaussian component produced an improvement in the fit ($\Delta \mathrm{C}_{sim}$), with respect to the continuum model, greater than the one we observed in the real data ($\Delta \mathrm{C}_{real}$). This translates into the probability that our observed feature is actually just a count fluctuation.
    The significance of the line detection is therefore:
    \begin{equation}
        P = 1 - \frac{N_{[\Delta \mathrm{C}_{sim} > \Delta \mathrm{C}_{real}]}}{N_{tot}}.
    \end{equation}

\begin{table}[t]
\def\arraystretch{1.17}
\begin{center}
\newcolumntype{R}{>{\raggedleft\arraybackslash}X}
\newcolumntype{L}{>{\raggedright\arraybackslash}X}
\begin{tabularx}{\columnwidth}{LLc}
\toprule
\textbf{Parameter} & \textbf{Value} & \textbf{Units} \\ \midrule
log$(\xi)$ & 3.69 $\pm$ 0.08 & - \\
N$_{\mathrm{H}}$ & $(2.3 \pm 0.9) \times 10^{23}$ & cm$^{-2}$ \\
V$_{\mathrm{OUT}}$ & 0.16 $\pm$ 0.02 & c \\
\bottomrule
\end{tabularx}
\end{center}
\caption{XSTAR table parameter values. Outflow velocity V$_{\mathrm{OUT}}$ was obtained from the $z$ parameter.}
\label{tab:UFO}
\end{table}

For the 2005 and 2020 observations we first tested the inclusion of a narrow line. We also tested a broad line for the 2020 observations, with $\sigma$ set to 0.47 keV, and constrained between 0.2 and 0.8 keV, motivated by previous results. 
From our simulations, none of the narrow detections (2 in 2005 and 2 in 2020) are significant ($\lesssim1\upsigma$).
Instead, the broad 2020 8.2 keV feature is significant at $98\%$.
We note that this is larger than the threshold of significant detection commonly used in population studies (95$\%$, \citealp{tombesi2010evidence,matzeu2023supermassive}). 

This result motivates us to further analyze the 2020 data. Instead of one broad Gaussian, we fit ad hoc XSTAR tables (\citealp{kallman1999xstar}). These models can be used to compute the physical conditions of the outflow, which are parametrized by the column density (N$_\mathrm{H}$), the ionization parameter ($\xi$), and the turbulent velocity ($v_{turb}$). The model also includes a redshift parameter $z$, which can be converted into an outflow velocity\footnote{The value of $z$ computed by the model is related to the absorber’s redshift ($z_a$) relative to the source’s
position ($z_c$) by $(1 + z) = (1 + z_a)(1 + z_c )$. This can be then translated into an outflowing velocity using the relativistic redshift formula $(1 + z_a)=\sqrt{(1-\mathrm{V_{OUT}}/c)/(1+\mathrm{V_{OUT}}/c)}$}.

\begin{table*}[t]
\def\arraystretch{1.17}
\begin{centering}
\newcolumntype{R}{>{\raggedleft\arraybackslash}X}
\newcolumntype{L}{>{\raggedright\arraybackslash}X}
\newcolumntype{C}{>{\centering\arraybackslash}X} % New column type for centered columns
\begin{tabularx}{\textwidth}{LCCCR}
\toprule
\textbf{Outflow Phase} & $\mathrm{V_{OUT}}$ & $\mathrm{\dot{M}_{OF}}$ $[\mathrm{M_{\odot}/yr}]$ & \multicolumn{1}{c}{$\mathrm{\dot{P}}$ $[L_{bol}/c]$} & \multicolumn{1}{c}{$\mathrm{\dot{K}}$ {[}erg/s{]}} \\ 
\midrule\textbf{UFO} & $0.16\pm0.02 \ c$ & $(7.5\pm4.2)\times 10^{-2}$ & $(3.1 \pm 1.7) \times  10^{-1}$ &   \multicolumn{1}{c}{$(5.5 \pm 3.1) \times 10^{43}$} \\
\textbf{Ionized ([OIII])} & 700$\pm$30 km/s & $3.7^{+0.5}_{-0.1}$ & $2.1^{+0.3}_{-0.6} \times 10^{-1}$ &  \multicolumn{1}{c}{ $5.6^{+0.8}_{-0.2} \times 10^{41}$} \\
\textbf{Ionized (H$\alpha$)} & 700$\pm$30 km/s & $17.4^{+0.9}_{-1.9}$ & $1.0\pm0.1$ &  \multicolumn{1}{c}{$2.6^{+0.1}_{-0.3} \times 10^{42}$} \\
\bottomrule
\end{tabularx}

\caption{Comparison of the parameters of disk-scale (UFO) and galaxy-scale (Ionized) winds. The parameters are also reported in the text and are respectively the outflow velocity, the mass outflow rate, the momentum outflow rate, and the kinetic power.}
\label{tabmulwind}
\end{centering}
\end{table*}
% \usepackage{booktabs}

\begin{figure}[t]
    \centering
    \includegraphics[width=\columnwidth]{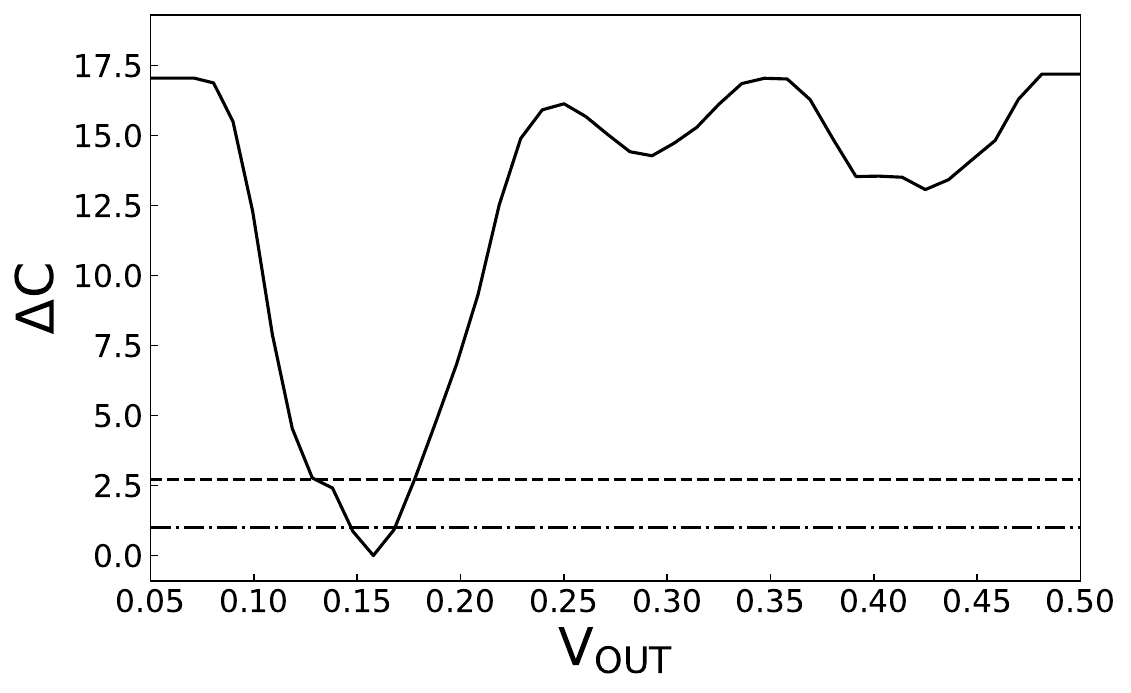}
    \caption{$\Delta\mathrm{C}$ value of the fit, as a function of the parameter z in UFO velocity space. The dash-dotted and dashed horizontal lines correspond respectively to 68\% and 90\% confidence regions.
}
    \label{fig:zcont}
\end{figure}

We tested two tables, one with a turbulent velocity of 1000 km/s and one with 5000 km/s. The data is best reproduced with 5000 km/s: after 
stepping the redshift parameter between -0.4 and 0.1, as first done in \cite{tombesi2011evidence}, we find a minimum at $z=-0.017 \pm 
0.008$, with a fit improvement of $\Delta\mathrm{C}$/d.o.f.=18/3 (Fig. \ref{fig:zcont}). The values of the XSTAR parameters are reported in Table \ref{tab:UFO}, and are fully consistent with typical Seyfert 1 UFOs.

We note that the fit improves much more significantly when the UFO is modeled as photoionized gas (XSTAR), rather than as a Gaussian absorption line. This is because the photoionized gas model accounts for a variety of absorption lines on a broad spectral range ($\sim$ 1 - 10 keV). As the outflow does not only contain iron, this model better reproduces the data, compared to a Gaussian absorption line, which only models the most prominent iron absorption feature (see e.g. \citealp{pounds2006confirming}).

\subsection{UFO energetics}

The UFO mass outflow rate can be derived as $\mathrm{\dot{M} = \Omega N_{H}}m_p\mathrm{V_{OUT}R}$, where $m_p$ is the proton mass, and $\Omega$ is the solid angle subtended by the outflow. We adopt the same recipe as in \cite{nardini2018multi}: $\Omega/4\pi = 0.5$ and R=$2c^{2}/\mathrm{V_{OUT}^2}$, 
which is the escape radius in units of gravitational radii $r_g = GM_{BH}/c^2$. This assumes that the UFO has reached a terminal velocity equivalent to this escape velocity at a distance R from the SMBH. Additionally, as discussed in \cite{luminari2020importance}, 
the relativistic velocities of UFOs can cause the underestimation of the outflow column densities by a factor 20-120\% for $\mathrm{V_{OUT}=0.1c-0.4c}$, as a result of beaming effects. This can significantly offset the value of $\mathrm{\dot{M}}$. The intrinsic UFO column density is larger than the observed XSTAR $\mathrm{N_H}$ by a factor $\Psi = (1+\beta)/(1-\beta) = 1.38\pm0.06$, where $\beta= \mathrm{V_{OUT}}/c$. 
This corresponds to a mass outflow rate $\mathrm{\dot{M} = (7.5\pm4.2) \times 10^{-2} \ M_{\odot}/yr}$, a momentum outflow rate $\mathrm{\dot{P} = (3.1 \pm 1.7) \times 10^{-1}} L_{bol}/c$ and $\mathrm{\dot{K} = (5.5 \pm 3.1) \times 10^{43} \ erg/s}$. These values are reported in Tab. \ref{tabmulwind}, together with the energetics of the galaxy-scale ionized outflow derived from [OIII] and H$_\alpha$.

\section{X-ray spectral analysis II: Populating the Lightcurve}
\label{sec:xrayII}

In this Section, we derive the 20-year-long X-ray lightcurve in the 0.5-2 keV band, using the observations listed in Tab. \ref{tab:lc}. The energy range was chosen both for its sensitivity to variations in N$_\mathrm{H}$, and because it guarantees good performance for all of the instruments considered in this analysis.
The 2005 and 2020 spectra have already been presented and analyzed. In the following subsections, we analyze each observation in chronological order. \textcolor{black}{The results of our fit procedure are shown in Table \ref{tab:lcfit}.}
\subsection{\textit{Chandra} - 2001}

\begin{figure}[t]
    \centering
    \includegraphics[width=\columnwidth]{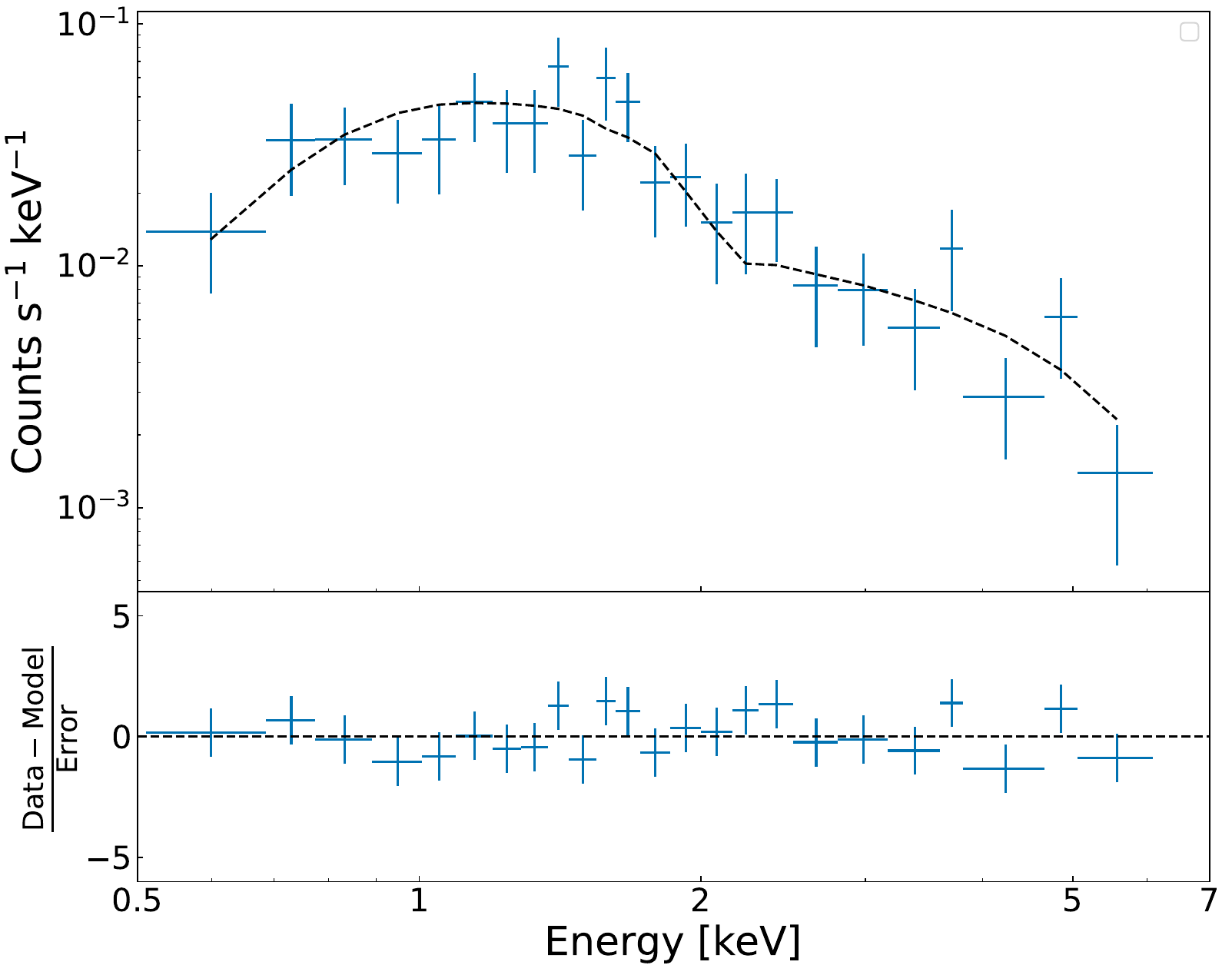}
    \caption{Spectrum (top) and residuals (bottom) of the absorbed power-law model of the 2001 \textit{Chandra} observation of 2M0918.}
    \label{fig:chandra}
\end{figure}

The \textit{Chandra} spectrum has a total of 155 counts in the 0.5-7 keV range. Prompted by the previously obtained results, we model the 2001 spectrum in \textsc{xspec} with an absorbed power-law model (\textsc{tbabs*ztbabs*powerlaw}). The data are reproduced nicely (Fig. \ref{fig:chandra}), with a C-stat value of 150 over 178 d.o.f.

\begin{figure}[t]
    \centering
    \includegraphics[width=\columnwidth]{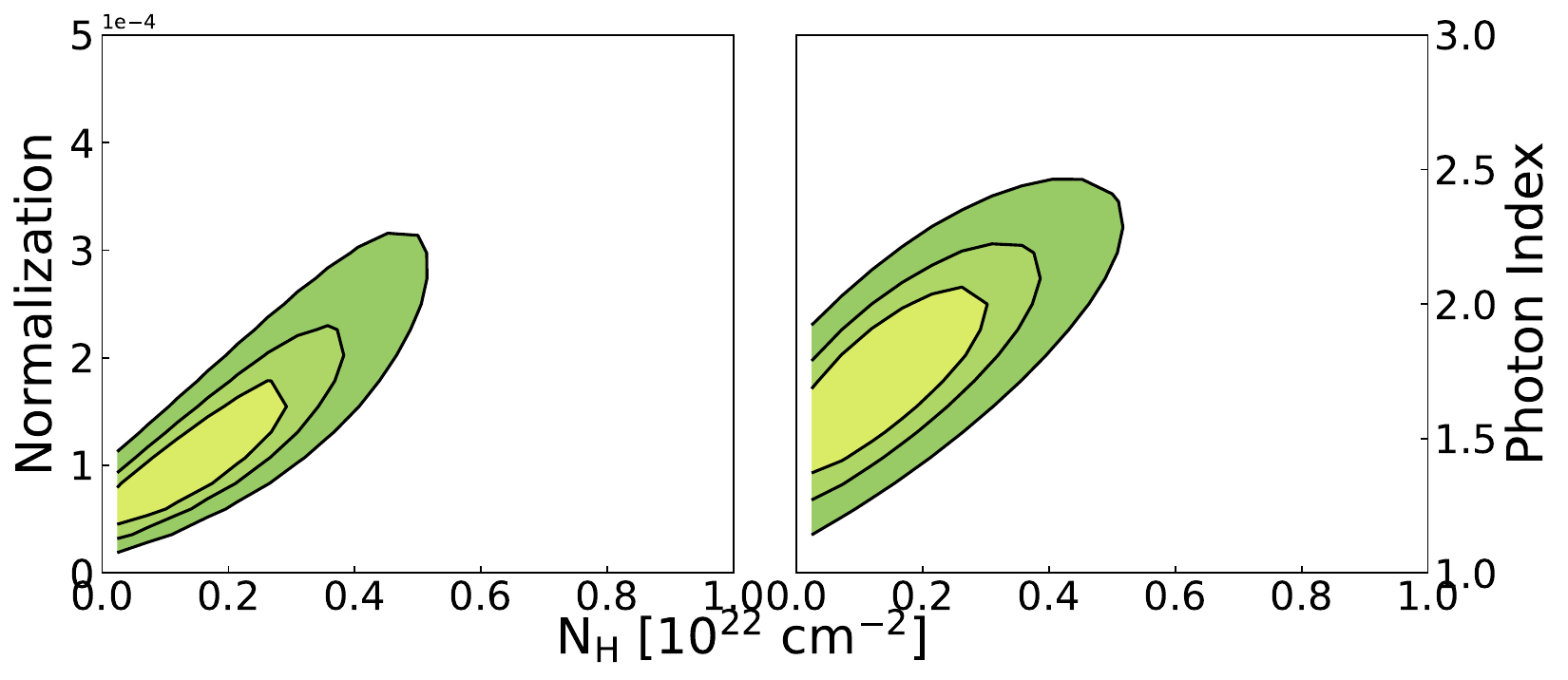}
    \caption{Contour plots of the column density of the absorber (x-axis) and the normalization (left) /photon index (right) of the primary power-law for the \textit{XMM-Newton} 2001 observation. In all cases, the observations are consistent with the source being unobscured.}
    \label{fig:chandracont}
\end{figure}

We obtain a photon index for the power-law of $\Gamma=1.8 \pm 0.4$ with normalization of $(1.75^{+0.9}_{-0.6}) \times 10^{-4} $ photons/keV/cm$^{2}$/s, while the absorber can only be constrained as an upper limit of N$_\mathrm{H}=3.6 \times 10 ^{21}$ cm$^{-2}$ at 90\% confidence level. This is also evident from the contour plots in Fig. \ref{fig:chandracont}, where the large uncertainties in parameter estimation are clear. 

We note that the photon index is consistent with the one derived for the 2005 and 2020 observations, while the power-law normalization seems to be consistent only with the 2020 spectra.

\subsection{\textit{XMM-Newton} - 2003}
\label{sec:xmm-2003}
The 2003 spectra are comprised of 467, 208
and 153 counts for pn, MOS1, and MOS2 respectively.
We initially model the three spectra simultaneously in the 0.5-10 keV range with the same simple absorbed power-law model as it was done in the \textit{XMM-Newton}
observation. However the fit is of poor quality (C-stat: 287/198 d.o.f.) and while the absorber is not needed in the fit, the inferred photon index is $\Upgamma=1.17\pm0.12$, which is well below the average AGN photon index of 1.8 (\citealp{nandra1994ginga,2008A&A...485..417D,ricci2017bat}). A photon index of less than 1.4 at 90\% confidence has been
used in literature as a selection criterion for highly obscured AGN (\citealp{lanzuisi2013chandra, georgantopoulos2013xmm, lanzuisi2018chandra}), as an absorbed power-law with an
intrinsically steeper photon index can be mimicked by an unabsorbed power-law with
a flatter photon index (\citealp{george1991x}). We, therefore, insisted on including an absorber in the model.

\begin{figure}[t]
    \centering
    \includegraphics[width=\columnwidth]{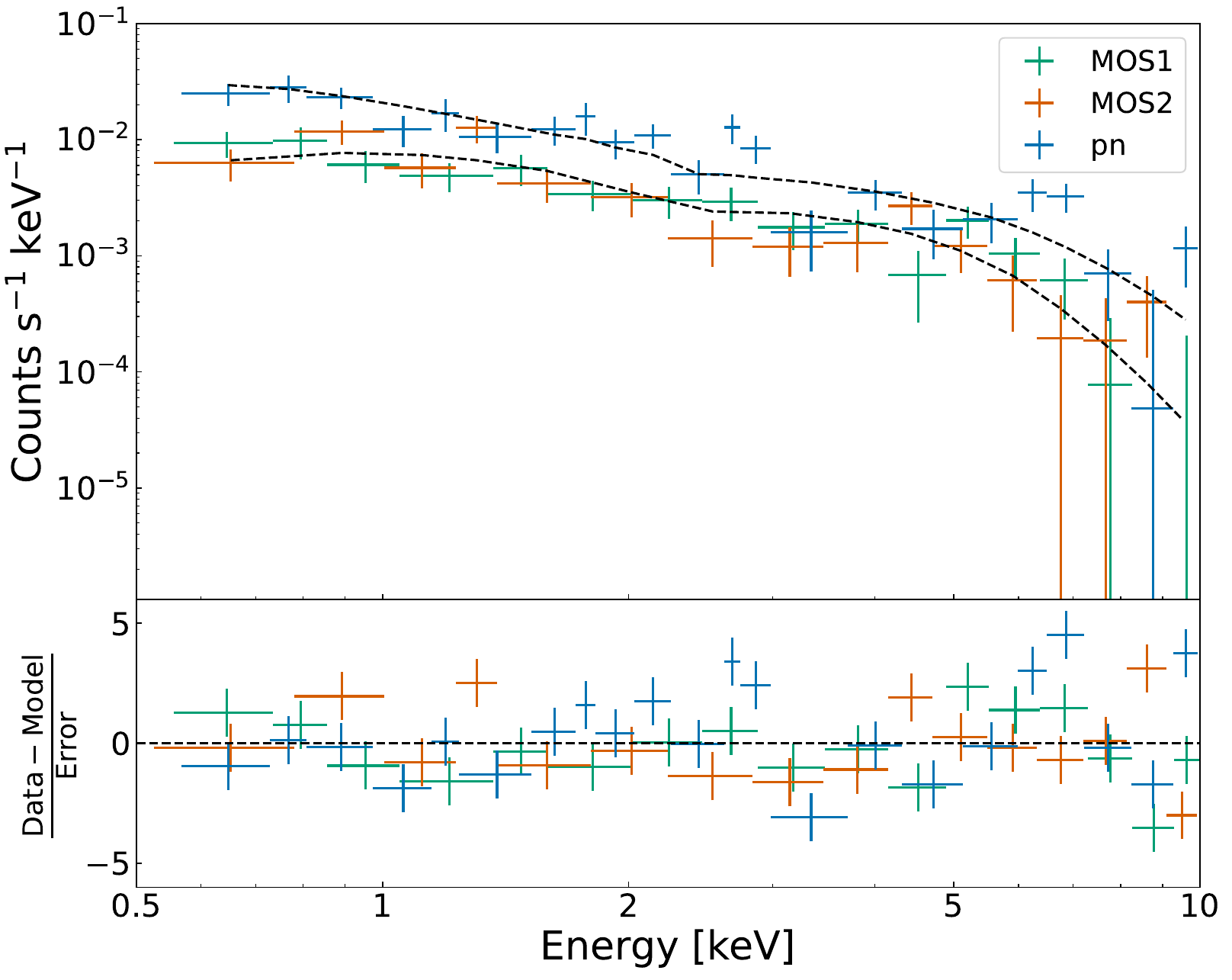}
    \caption{Spectrum (top) and residuals (bottom) of the partially absorbed power-law model for the 2003 \textit{XMM-Newton} observation of 2M0918.}
    \label{fig:spec2003}
\end{figure}

The previously used \textsc{zxipcf} model has too many parameters, for which degeneracies are not solvable with such photon statistics. Given the evidence for the absorber being cold in the 2005 data, we opted for the cold partial covering absorber model \textsc{TBPCF}. With a C-stat value of 280/196 d.o.f, we find a covering fraction of 0.76$^{+0.1}_{-0.17}$, and a photon index $\Gamma = 2.03 \pm 0.14$, both in agreement with the results obtained in Sect. \ref{sec:xrayI}. The partially covering absorber has a column density of N$_{\mathrm{H}}$ = $6.4^{+4.2}_{-2.0} \times 10^{22}$ cm$^{-2}$.

We note that in the hard band some spectral features, such as iron lines, can still
be present in the residuals in Fig. \ref{fig:spec2003}, but as the focus of our spectral analysis is soft
flux estimation we are satisfied with our best-fit model.

\begin{figure*}[t]
    \includegraphics[width=\textwidth]{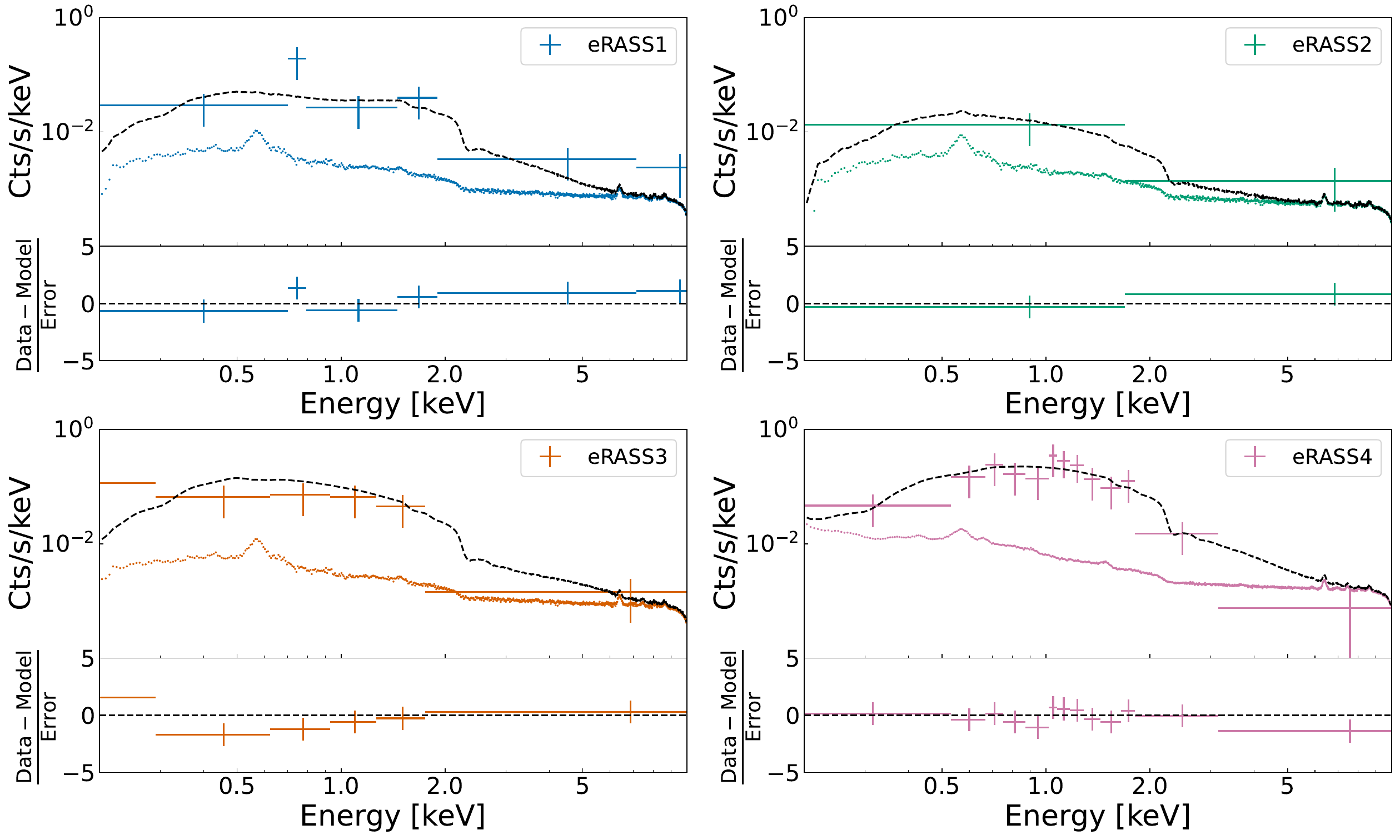}
    \caption{Spectra and residuals of the 4 eROSITA observations. The spectra are binned for visualization purposes. In the same color as the spectra, the PCA background model is shown for each observation, while the dashed lines show the best-fit model for each eRASS. Note the drastic photon drop in eRASS2 and the rise in eRASS4.}
    \label{Fig:erospec}

\end{figure*}

\subsection{eROSITA - 2020-2021}

Given the low exposure times (see Table \ref{tab:lc}), and the low photon counts (respectively 17, 5, 17, and 39 photons for eRASS1-4), we analyzed the eROSITA spectra within the Bayesian framework. We do so with the software \textsc{Bayesian X-ray Analysis} (BXA\footnote{\url{https://johannesbuchner.github.io/BXA/}}, \citealp{buchner2014x}) which connects the nested sampling algorithm UltraNest\footnote{\url{https://johannesbuchner.github.io/UltraNest/}} package (\citealp{2021JOSS....6.3001B}),
with the fitting environment CIAO/Sherpa (\citealp{freeman2001sherpa,fruscione2006ciao}).

We chose to use this analysis technique as it is particularly well suited for low photon statistics, as no binning or assumption of gaussianity of the parameter distribution is needed. Moreover, BXA also includes the possibility to simultaneously model the background with empirical Principal Component Analysis (PCA) models (see \citealp{simmonds2018xz} for details on background PCA models).

We model the four unbinned spectra and backgrounds (shown in Fig. \ref{Fig:erospec}) simultaneously in the 0.2 - 10 keV range. \textsc{Sherpa} includes \textsc{xspec} models, therefore for consistency with all of the \textit{XMM-Newton}(+\textit{NuSTAR}) observations, we model the source as a power-law absorbed by \textsc{tbpcf} with covering fraction fixed at 80\%, with the usual Galactic absorption. BXA, being of Bayesian nature, allows us to include priors on the parameters, which we set to be uniform on the log of the power-law normalization and the column density of the absorber. This means that we are assuming total ignorance of these parameters \textit{prior} to our analysis. We instead implement a Gaussian prior for the photon index with $\mu_\Gamma = 1.95$ and $\sigma = 0.15$, based on the thoroughly-observed distribution discussed in Sec. \ref{sec:xmm-2003}. Given our low photon counts, and the softness of the eROSITA response, it is crucial to have some constraints on the photon index. 

\begin{figure}[t]
    \centering
    \includegraphics[width=\columnwidth]{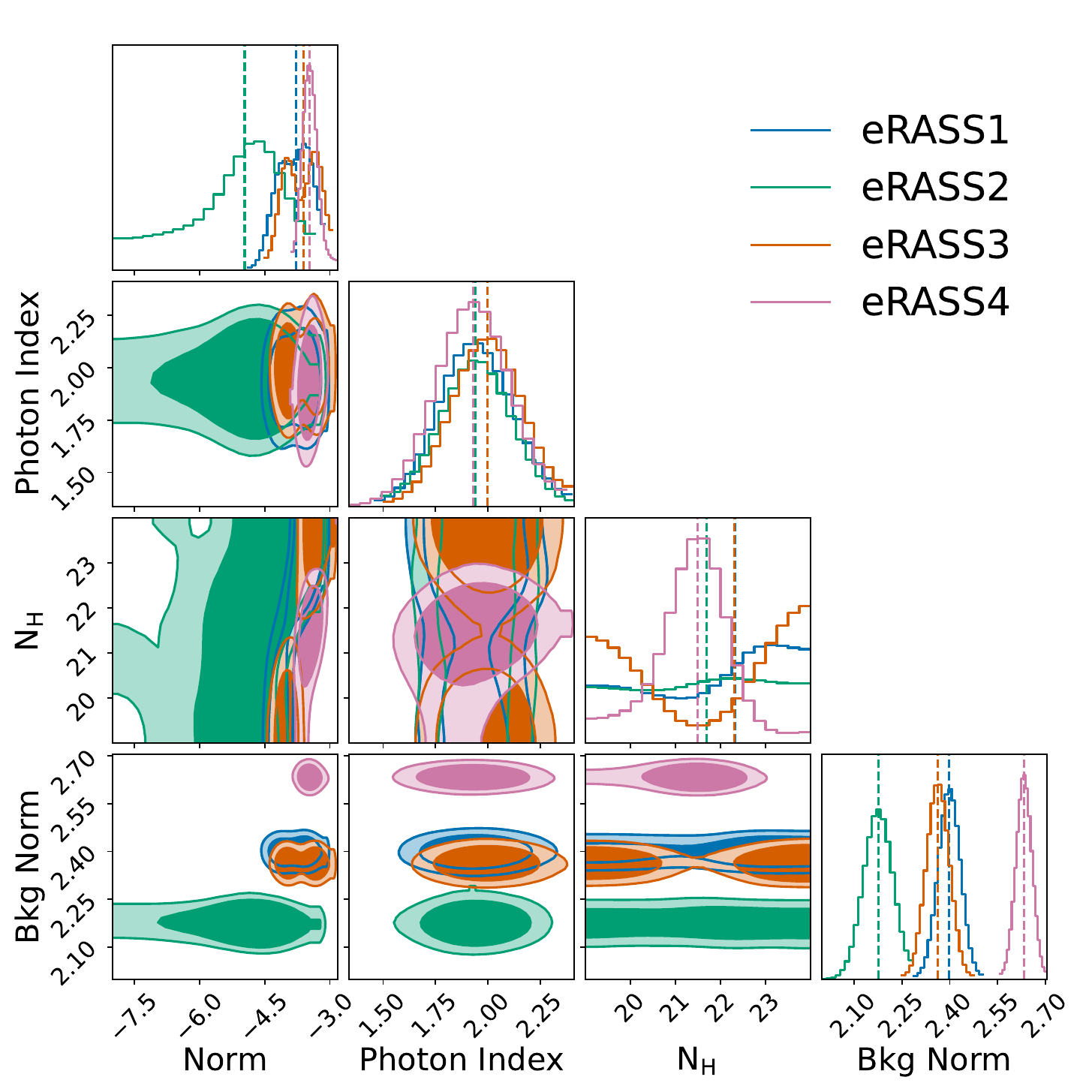}
    \caption{Cornerplot of the fitted parameters (power-law normalization, photon index,
column density and background normalization) for the eROSITA observations. The histograms represent individual parameter posterior distribution, while the shaded contours show the credibility regions of the interception of different parameters. The darker color corresponds to 68\% of the integrated posterior, while the lighter shades to 90\%. Color coding is as in Fig.\ref{Fig:erospec}.}
    \label{fig:corner}
\end{figure}

Figure \ref{fig:corner} shows the resulting corner plot, produced with the Python module Corner.py (\citealp{corner}), where contours represent credibility regions, rather than confidence levels\footnote{A xx\% credibility region is defined as the region encompassing xx\% of the posterior distribution.}. 
The fit is insensitive to the local absorber in all eRASSes except 4 (the brightest), in which we find moderate absorption \textcolor{black}{($\log(\mathrm{N_{H}}) = 21.49^{+0.74}_{-1.73}$)}. We also note that the source is consistent with being undetected in eRASS2.

\subsection{The 20-year-long lightcurve}

\begin{table*}[t]
\textcolor{black}{
\def\arraystretch{1.17}
\begin{center}
\newcolumntype{R}{>{\raggedleft\arraybackslash}X}
\newcolumntype{L}{>{\raggedright\arraybackslash}X}
\begin{tabularx}{1\textwidth}{ccLLc}
\toprule
\textbf{Observation}              & \textbf{Component}                 & \textbf{Parameter}                 & \textbf{Value}                       & \textbf{Units}      \\ \midrule
\textbf{\textit{Chandra} - 2001} & \textsc{ztbabs}   & N$_{\mathrm{H}}$                   & $<3.6 \times 10^{21}$         & cm$^{-2}$          \\
\textbf{}                         & \textsc{powerlaw} & Photon index                       & $1.8\pm0.4$                        & \textbf{-}          \\
\textbf{}                         & \textbf{}                          & Normalization                      & $(1.75^{+0.9}_{-0.6} \times 10^{-4}$         & photon/keV/cm$^{2}$/s \\ \midrule
\textbf{\textit{XMM} - 2003} & \textsc{tbpcf}   & N$_{\mathrm{H}}$                   & $6.4^{+4.2}_{-2.0} \times 10^{22}$ & cm$^{-2}$           \\
                                  &                                    & Covering fraction                  & $0.76^{+0.1}_{-0.17}$               & -                   \\
                                  & \textsc{powerlaw} & Photon index & $2.03\pm 0.14$               & -                   \\
                                  &                                    & Normalization                      & $(1.4\pm0.2) \times 10^{-4}$         & photon/keV/cm$^{2}$/s \\ \midrule
\textbf{eRASS1} & \textsc{tbpcf}   & N$_{\mathrm{H}}$                   & - & cm$^{-2}$           \\
                                  &                                    & Covering fraction                  & 0.8               & -                   \\
                                  & \textsc{powerlaw} & Photon index & $1.94\pm0.24$               & -                   \\
                                  &                                    & Normalization                      & $1.6^{+2.8}_{-1.1} \times 10^{-4}$         & photon/keV/cm$^{2}$/s \\ \midrule
\textbf{eRASS2} & \textsc{tbpcf}   & N$_{\mathrm{H}}$                   & - & cm$^{-2}$           \\
                                  &                                    & Covering fraction                  & 0.8              & -                   \\
                                  & \textsc{powerlaw} & Photon index & $1.94^{+0.23}_{-0.25}$               & -                   \\
                                  &                                    & Normalization                      & $<1 \ \times 10^{-4}$         & photon/keV/cm$^{2}$/s \\ \midrule
\textbf{eRASS3} & \textsc{tbpcf}   & N$_{\mathrm{H}}$                   & - & cm$^{-2}$           \\
                                  &                                    & Covering fraction                  & 0.8              & -                   \\
                                  & \textsc{powerlaw} & Photon index & $2.00^{+0.23}_{-0.25}$               & -                   \\
                                  &                                    & Normalization                      & $2.4^{+4.4}_{-1.7} \times 10^{-4}$         & photon/keV/cm$^{2}$/s \\ \midrule
\textbf{eRASS4} & \textsc{tbpcf}   & N$_{\mathrm{H}}$                   & $3.1^{+14.6}_{-3.0} \times 10^{21}$ & cm$^{-2}$           \\
                                  &                                    & Covering fraction                  & 0.8              & -                   \\
                                  & \textsc{powerlaw} & Photon index & $2.03^{+0.24}_{-0.24}$               & -                   \\
                                  &                                    & Normalization                      &  $3.4^{+3.4}_{-1.3} \times 10^{-4}$         & photon/keV/cm$^{2}$/s \\ \midrule
\end{tabularx}
\end{center}
\caption{Values for best-fit continuum models for the six observations analyzed to build the light curve. The empty column density entries reflect the fact that the fit could not constrain the parameter.}
\label{tab:lcfit}
}
\end{table*}

\begin{figure*}
    \includegraphics[width=\textwidth]{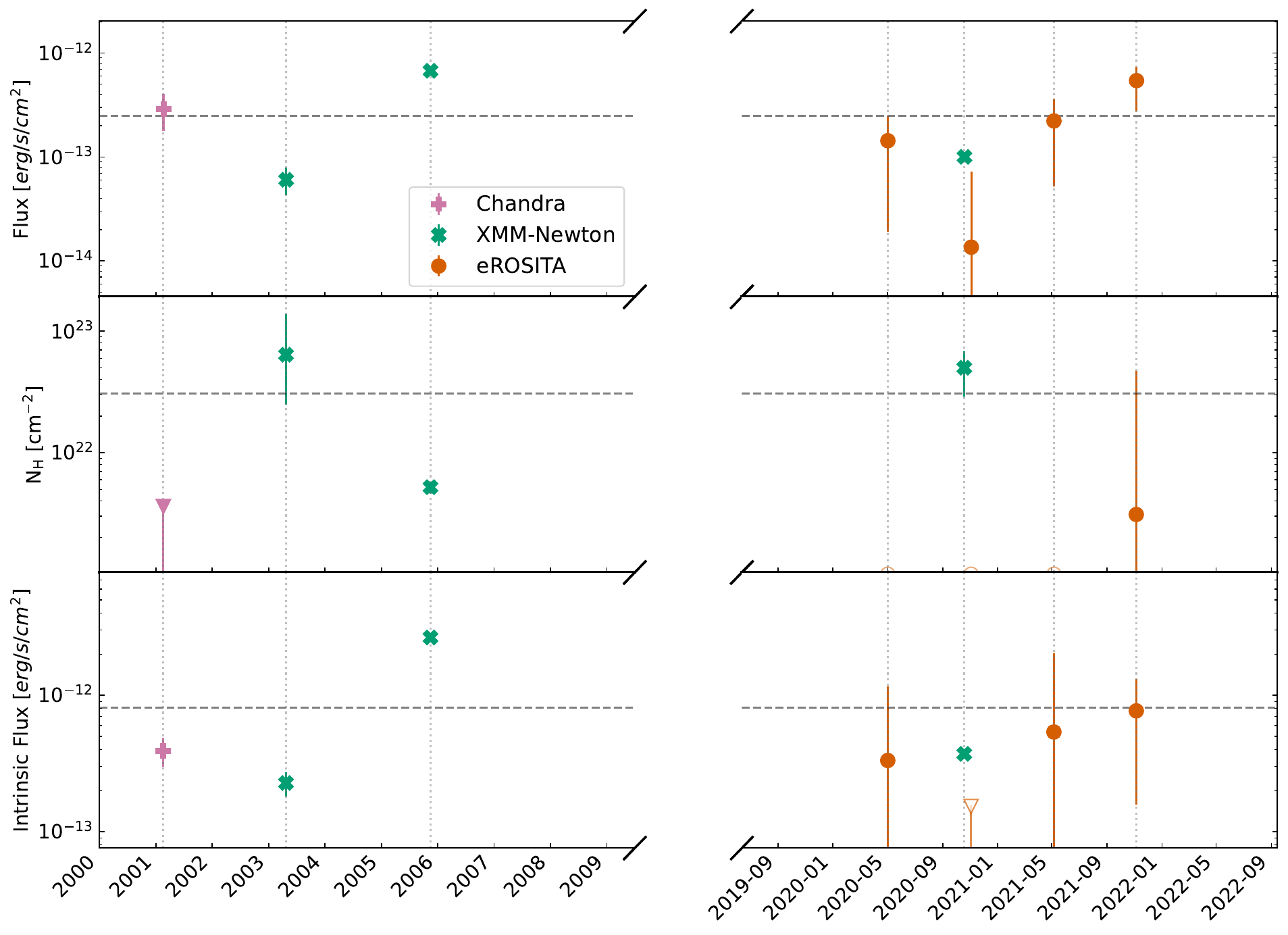}
    \caption{\textit{Top}: 0.5-2 keV observed-flux lightcurve of 2M0918. In all panels, the dashed horizontal line represents the weighted average of the y values. \textit{Middle:}  Absorber $\mathrm{N_H}$ over time. The red hollow
circles correspond to eRASS 1-3, where the fit did not constrain the column density. \textit{XMM-Newton} is reported as an upper limit at 90\% confidence. A clear anticorrelation with flux is noticeable when $\mathrm{N_H}$ is compared with the top panel. \textit{Bottom}: Intrinsic flux light-curve. eRASS2 was only constrained as an upper limit.}
    \label{fig:lc}
\end{figure*}
We show the 20-year-long observed flux X-ray lightcurve in Fig. \ref{fig:lc} (top), which includes fluxes from all the presented spectra.
Variability is evident and recurrent, with the 2005 \textit{XMM-Newton} observation being 11 times brighter than the 2003 observation and 6 times brighter than the 2020 observation. The dimmest state is the one associated with eRASS2, where the source was consistent with being undetected. However, this was only 14 days after a clear \textit{XMM-Newton} detection, and if we take the upper error bar to be the 3$\sigma$ upper limit on the eRASS2 flux, the 2 values are compatible with small variability.
The flux did however increase again by a factor 5.3 with respect to the 2020 \textit{XMM-Newton} observation in eRASS4.

The middle panel of Fig. \ref{fig:lc} shows the value of N$\mathrm{_{H}}$ throughout the epochs, in clear anticorrelation with the observed flux (we exclude eRASS 1-3 as the fit was insensitive to variations in column density for these observations). In other words, dimmer states correspond to higher inferred column density, while brighter states correspond to lower quantities of obscuring material. This would favor a scenario in which the recurring alternating between high and low flux states is driven by a change in the absorbing material, which is not uncommon (see e.g. \citealp{torricelli2014search} and references therein for other examples of variable N$\mathrm{_{H}}$ in AGN).

We also plot the intrinsic flux variation over time the bottom panel of the same figure. 
As one can see, the intrinsic flux increases significantly between 2003 and 2005 and then drops again in 2020. The flux ratios are 11.6 (2005/2003) and 7 (2005/2020). 

We note that the eRASS2 upper limit is not consistent with the 2020 \textit{XMM-Newton} intrinsic flux measurement. The data seem to indicate a drop of a factor $\sim2$ in only 14 days.
 
\section{Discussion}
\label{sec:disc}
In this section, we first present a discussion on the multiphase and multiscale winds, and subsequently a discussion on the interpretation of the observed variability. We finalize with a discussion on future perspectives of this work.
\subsection{Multiphase and multiscale outflows}

One of the most relevant features of our analysis is the simultaneous detection of multiphase winds. As mentioned in the introduction, AGN feedback is thought to propagate from small, sub-pc scales all the way to galaxy and group/cluster scales. Although this is commonly accepted both from a theoretical and computational point of view (\citealp{king2015powerful}, \citealp{costa2020powering}), the physics of feedback is still not fully clear. More specifically, it is yet not known whether winds propagate in a momentum-driven fashion, in which radiation cooling is faster than the outflow, and so energy is not conserved, or in an energy-driven scenario, in which cooling is negligible and the wind shock-front expands adiabatically. While theory deems both mechanisms to be suitable, observations can constrain this by measuring disk- (UFOs) and galaxy-scale winds (molecular, atomic, or ionized phase) in the same galaxy. A collection of sources for which energetics in at least two phases are available has been compiled by \cite{marasco2020galaxy}, and later extended by 
\cite{tozzi2021connecting} and 
\cite{bonanomi2023another}. However, the sample is still small (17 sources + Mrk 509, reported in \citealp{zanchettin2021ibisco}), especially for what concern sources with simultaneous detection of UFOs and galaxy scale outflows in the ionized gas phase.
With 2M0918 we are able to expand the sample of sources in which both a UFO and an ionized outflow are detected from 5 to 6.

\begin{figure*}[h]
    \centering
    \includegraphics[width=\textwidth]{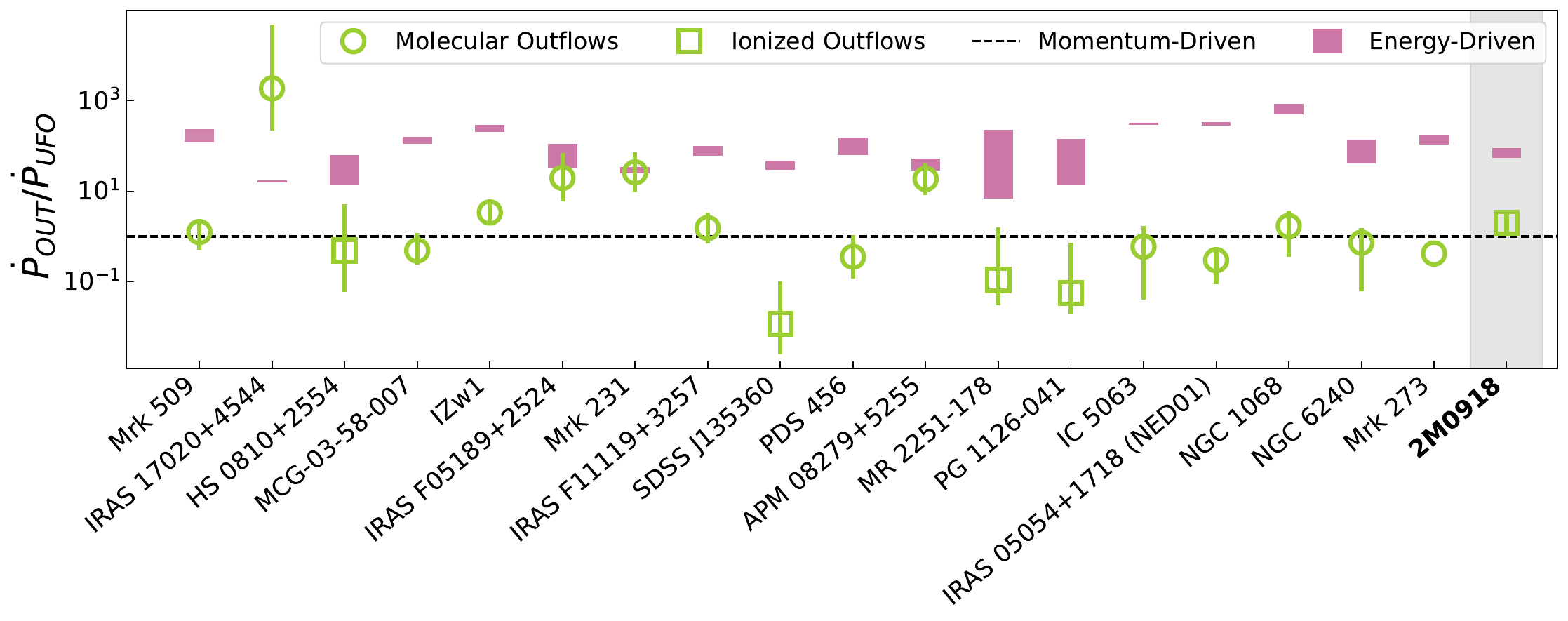}
    \caption{Updated figure from \cite{bonanomi2023another}, first presented in \cite{marasco2020galaxy} and \cite{tozzi2021connecting}, with the addition of 2M0918 (grey shaded area) and Mrk 509 (\citealp{zanchettin2021ibisco}). This plot contains a compilation of
sources for which UFO and galaxy-scale wind energetics are available, with the large-scale winds being either ionized or molecular/atomic. The ratio of the outflow momentum rates $\mathrm{\dot{P}_{OUT}/\dot{P}_{UFO}}$ is plotted for each source, together with values predicted from theory for momentum- or energy-driven scenarios. The predicted value can be estimated as the ratio of $\mathrm{V_{OUT}/V_{UFO}}$ by imposing the
conservation of energy, while the dashed line at $\mathrm{\dot{P}_{OUT}/\dot{P}_{UFO}}$ =1 is the prediction of a momentum-driven scenario. For 2M0198, we use the average between the $\mathrm{\dot{P}_{OUT}}$ derived from [OIII] and H$_\alpha$. \textcolor{black}{The compilation uses values reported in \cite{bischetti2019gentle, braito2018new, chartas2009confirmation, cicone2014massive, feruglio2015multi, feruglio2017discovery, garcia2014molecular, gonzalez2017feedback, longinotti2015x, longinotti2018early, lutz2020molecular, marasco2020galaxy, mizumoto2019kinetic, reeves2019momentum, rupke2017quasar, sirressi2019testing, smith2019discovery, tombesi2015wind, tozzi2021connecting, veilleux2017quasar,zanchettin2021ibisco}}}
    \label{fig:tozzi}
\end{figure*}

In Fig. \ref{fig:tozzi}, we add \textcolor{black}{2M0918 and Mrk 509 to the diagnostic plot presented in \cite{marasco2020galaxy}, \cite{tozzi2021connecting} and \cite{bonanomi2023another}}. The vertical axis shows the ratio of the momentum outflow rates, respectively of the galaxy- and disc-scale winds. Observations consistent with a momentum-driven scenario would lie on the horizontal dashed line, while the pink squares represent the energy-driven predictions. As shown in the grey-shaded area, 2M0918 is consistent with a momentum-driven scenario.
Our results corroborate the conclusion drawn by \cite{marasco2020galaxy} and \cite{tozzi2021connecting}, which state that the energy- and momentum-driven models of feedback propagation explain well the observation at least up to galaxy scales. 
Note that for some of the sources, the points fall even below the momentum-driven prediction. This can be likely explained by the fact that ionized winds are ejected hundreds to thousands of years before the observations, therefore, X-rays might be tracing a more energetic, later event.

\begin{figure}[h]
    \centering
    \includegraphics[width=\columnwidth]{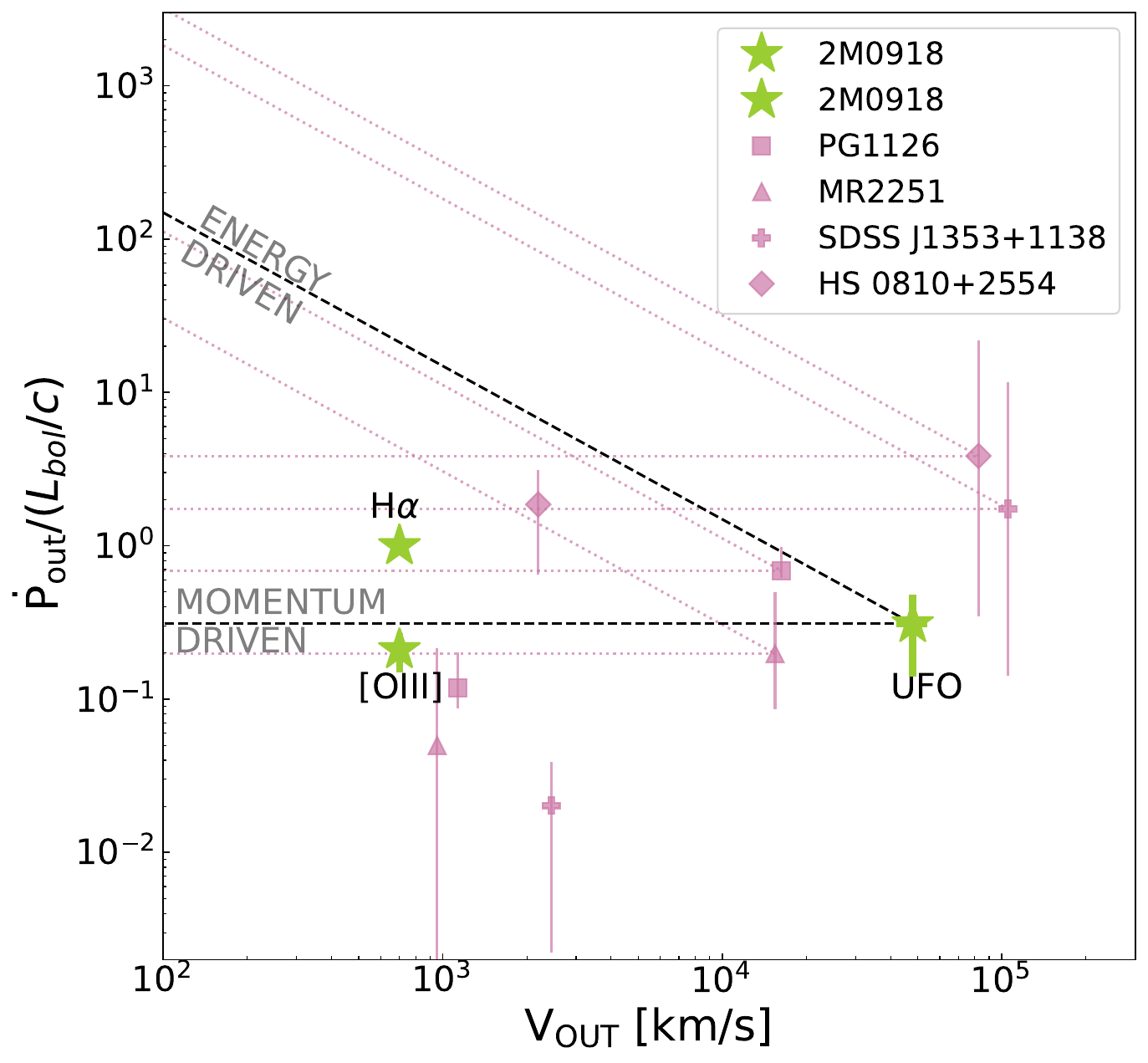}
    \caption{Momentum outflow rate normalized by $L_{bol}/c$ vs. outflow velocity for 2M0918 (green stars, different components as labeled). The two solid lines correspond to the two different feedback mechanisms described in the text, as labeled. In the background, we also plot the sources from \cite{marasco2020galaxy} and \cite{tozzi2021connecting} where galaxy scale
winds are ionized, each with its respective diagnostic line.}
    \label{fig:enemom}
\end{figure}

In Fig. \ref{fig:enemom} we plot the outflow momentum rate of 2M0918 normalized by $L_{bol}/c$ as a function of outflow velocity (green stars), together with the other five sources with detected UFOs and ionised outflows (pink). This is another standard diagnostic for feedback mechanisms, as winds conserving momentum in their propagation would lie on the dashed line parallel to the x-axis, while energy-driven winds would follow the descending line. 
Once again, despite the energetics derived from the [OIII] and H$_\alpha$ lines differing by almost one order of magnitude, 2M0918 generally occupies a region of the plane consistent with a momentum-driven scenario.

%In Fig. \ref{fig:tozzi} we add 2M0918 (and Mrk 509, of which a multiphase wind analysis is presented in \citealp{zanchettin2021ibisco}) to the plot initially presented in \cite{marasco2020galaxy} extended with the \cite{bonanomi2023another} sources. While we already knew that our source is likely associated with a momentum-driven scenario, it is worth mentioning that our results corroborate the conclusion drawn by both of the authors, which state that the energy- and momentum-driven models of feedback propagation explain well the observation at least up to galaxy-scales. 
%Note that for some of the sources, the points fall below the momentum-driven prediction. This can be likely explained by the fact that ionized winds are ejected hundreds to thousands of years before the observations, therefore X-rays might be tracing a more energetic, later event.

It is important to mention that the radius inferred for the optical ionized outflow is technically an upper limit, as we are using the diameter of the SDSS optical fiber. However, typical outflow radii for the ionized phase are of the order of $\sim1$ kpc (See e.g. \citealp{fiore2017agn}), therefore we do not expect to be underestimating the momentum outflow rate by a factor larger than 4, which would not change the feedback propagation scenario. In addition to this, recent results (e.g. 
\citealp{baron2019discovering}, \citealp{davies2020ionized}), which made use of the \textit{Trans-Auroral Ratio} (TR) method first introduced in \cite{holt2011impact} to determine $n_{e_3}$ in outflows, suggest that the [SII] doublet method might systemically underestimate the true electron densities by even one order of magnitude. With our data, we cannot test this method, as it requires the observation of lines that would fall at infrared wavelengths. As an increase in the density of the gas would push the $\mathrm{\dot{P}_{out}}$ down by an equal factor, the interplay between this uncertainty and the uncertainty on the radius makes us confidently state that our observations are consistent with a momentum-driven scenario.

\subsection{Tracks in the $N_H$-$\lambda$ plane and bolometric Luminosities}

\begin{figure}{h}
    \centering
    \includegraphics[width=\columnwidth]{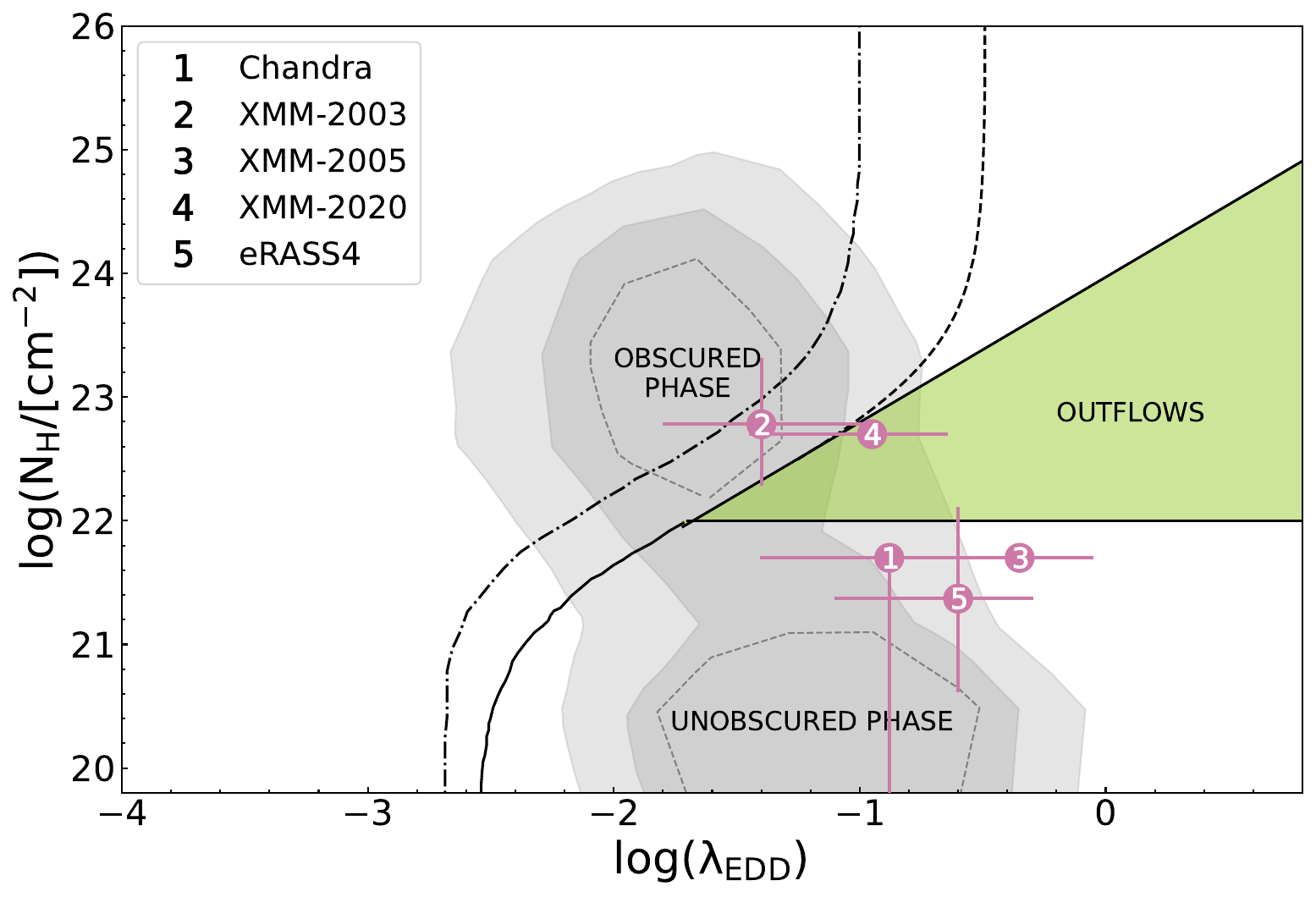}
    \caption{$\mathrm{N_H} - \lambda_{EDD}$ plane, adapted from \cite{ricci2022bass}, and including density
contours of the BASS XXXVII sample (grey areas). The green region labeled as “OUTFLOWS”
(also known in literature as forbidden region) corresponds to an $\mathrm{N_H} - \lambda_{EDD}$ space where
absorption cannot be long-lived. The \textcolor{black}{dot-dashed line corresponds to the effective Eddington limit for dusty gas reported in \cite{ishibashi2018energetics}, while the dashed line is
the effective Eddington limit when including infrared radiation trapping, from the same authors, adapted to the seminal values of \cite{fabian2009radiation} (solid black line
constraining the forbidden region), similarly to what was done by \cite{lansbury2020x}}. 2M0918 entered and left the outflow region twice between 2001 and 2021, in agreement with the detection of outflows.}
    \label{fig:nhlambda}
\end{figure}

AGN caught in active feedback phases are expected to be located in a specific region of the $\mathrm{N_{H}}-\lambda_{EDD}$ plane (\citealp{fabian2008effect}).
This is due to the interplay between the intensity of radiation pressure from accretion and the gravitational force that acts on the dusty obscuring material. In other words, as the accretion rate increases, only very heavy nuclear absorbing clouds can survive the intense radiation field. In Fig. \ref{fig:nhlambda} we plot the N$_{H}-\lambda_{EDD}$ plane, as adapted from \cite{ricci2022bass}, with the density contours found from the Swift-BAT AGN Spectroscopic Survey sample (BASS) presented in the same paper. The positions spanned by 2M0918 are included in the plot.

While the column density $\mathrm{N_H}$ was derived from the spectral analysis described previously, the Eddington ratio is defined as $\lambda_{EDD} = L_{bol}/L_{EDD}$, with the Eddington luminosity $L_{EDD} = 1.26 \times 10^{38} (M/M_{\odot})$ being a constant only dependent on the mass of the black hole 
(for a mass of $\mathrm{log(M/M_{\odot}) = 7.4\pm 0.4}$ this is $L_{EDD} = (3.2^{+4.8}_{-1.9}) \times 10^{45}$ erg/s). Even though SED fitting is generally considered one of the best probes for the bolometric luminosity, it is not a tracer of variability. By assuming that changes in the intrinsic X-ray emission are due to variations in accretion rate, deriving the bolometric luminosity from the 2-10 keV band is more suited to account for variability. 

For our purposes, therefore, we estimate the AGN bolometric luminosity using the bolometric correction ($k_{bol} \equiv L_{bol}/L_{2-10})$. We use the bolometric corrections presented in \cite{duras2020universal}:
\begin{equation}
    k_{bol(L_{2-10})} = 15.33 \times \bigg[ 1+ \bigg(\frac{log(L_{2-10}/L_{\odot})}{11.48} \bigg)^{16.2}\bigg].
\end{equation}

By applying this equation to the \textit{XMM-Newton}, \textit{XMM-Newton}(+\textit{NuSTAR}), and eROSITA-eRASS4 2-10 keV fluxes we obtain the $\lambda_{EDD}$ in Tab. \ref{tab:edd}.
From these results and referring to Fig. \ref{fig:nhlambda}, it is evident that 2M0918 has been living for the past 20 years in the vicinity of the ``forbidden" outflow region of the $\mathrm{N_{H}}-\lambda_{EDD}$ plane, possibly crossing it throughout its evolution. Moreover, the 2020 observation, in which UFOs were detected, falls inside the outflow region, although errors are large due to the uncertainty on the SMBH mass. Changes in accretion rate are also present, which provides us with further hints on the nature of the variability of 2M0918, which we describe and interpret in the next Section. Overall, these results are consistent with our conclusions and demonstrate the potential of using the $N_{H}-\lambda_{EDD}$ plane for short-time variability, rather than secular evolution.

We stress that the underlying assumption in this treatment is that winds are radiation-driven. It should be noted that other mechanisms, which we do not explore in this work, can also be responsible for the launching of winds. These include but are not limited to magnetically-driven winds, (e.g. \citealp{lynden1996magnetic, yuan2015numerical, fukumura2022tell}) and thermally-driven winds (e.g. \citealp{begelman1983compton, waters2018magnetothermal}).

\begin{table}[t]
\def\arraystretch{1.17}
\begin{centering}
\newcolumntype{R}{>{\raggedleft\arraybackslash}X}
\newcolumntype{L}{>{\raggedright\arraybackslash}X}
\newcolumntype{C}{>{\centering\arraybackslash}X}
\begin{tabularx}{\columnwidth}{LR}
\toprule
\textbf{Observation} & \textbf{$\lambda_{EDD}$} \\ \midrule
\textit{Chandra} - 2001 & 0.13$^{+0.20}_{-0.08}$ \\
\textit{XMM-Newton} -2003 & 0.05$^{+0.08}_{-0.03}$ \\
\textit{XMM-Newton} - 2005 & 0.44$^{+0.67}_{-0.26}$ \\
\textit{XMM}+\textit{NuSTAR} - 2020 & 0.11$^{+0.17}_{-0.07}$ \\
eRASS4 - 2021 & 0.25$^{+0.38}_{-0.15}$ \\ \bottomrule
\end{tabularx}
\caption{Eddington ratio for each observation. The large uncertainties on the mass strongly affect these values.}
\label{tab:edd}
\end{centering}
\end{table}

\subsection{The Nature of the Variability of 2M0918}

In Sect. \ref{sec:xrayII} we derived the 20-year-long X-ray lightcurve from cross-instrument spectral analysis. The choice of individually modeling each spectrum, instead of simply fitting all observations with the same power-law model and looking for changes in normalization, enables, with some degree of uncertainty, to disentangle the variability of different components. 

The spectral variability observed in 2M0918 is sufficiently drastic in order to classify this as a CL-AGN. 
\cite{ricci2022changing} divide CL-AGNs into two classes based on whether the variability is to be ascribed to changes in the line-of-sight obscuring material (\textit{changing-obscuration} AGN, CO-AGN \citealp{mereghetti2021time}) or to changes in accretion state (\textit{changing-state} AGN, CS-AGN, \citealp{graham2020understanding} ). 
X-ray CL-AGN are typically associated with CO-AGN, and some degree of obscuration variability is observable in a large fraction of the general AGN population (e.g. \citealp{risaliti2002ubiquitous}, \citealp{markowitz2014first}). This is usually attributed to eclipsing events from gas clouds in the BLR (this is the case for the famous NGC1365, \citealp{risaliti2009variable}, \citealp{maiolino2010comets}) or the clumpy nature of the dusty torus (see \citealp{almeida2017nuclear} for a review). There is no particular reason why we should not consider these as valid scenarios for the observed variability of 2M0918: timescales range from days to years and the variations in $\mathrm{N_{H}}$ are well within the observed range. However, these models do not predict or require any variation in the AGN accretion rate, which we observe (Fig. \ref{fig:nhlambda}). We would then have to explain the accretion changes as uncorrelated to the obscuration and the result of disk instabilities (\citealp{sniegowska2022modeling}). We stress once again that one should still keep these models in mind as possible explanations of the observed lightcurve, however, it is also possible to explore the plausibility of a connection between the two phenomena.

In the last decade, an alternative mechanism in which the CO-AGN event is explained in terms of obscuration due to outflowing gas material has been proposed for some sources (such as NGC5548, \citealp{kaastra2014fast}, NGC3227, \citealp{beuchert2015variable} and NGC378, \citealp{mehdipour2017chasing}). Additionally, \cite{marchesi2022compton} explained the X-ray spectral variability of NGC 1358 within a recurring feeding-feedback framework. In these cases, the absorber is ionized to some degree, but for low S/N observations, its imprint on the spectra can mimic that of neutral gas (see e.g. \citealp{waddell2023erosita}). 

For 2M0918 the ionization parameter $\xi$ was observed to increase by an order of magnitude between 2005 and 2020 (the two observation sets in which the S/N was high enough to disentangle the degeneracy in the \textsc{zxipcf} parameters), concurrent with an increase in column density. These are the same observations in which UFO signatures were detected.
At the same time, between 2003 and 2020, as it's shown in Fig. \ref{fig:nhlambda}, the AGN entered, crossed, and left the $\mathrm{N_H}-\lambda_{EDD}$ plane region in which sources are expected to be in the outflowing phase. We propose that, from these considerations, the observed variability can be associated with outflowing gas. 

We propose the following scenario, which explains the variability in accretion rate, absorber column density, and the appearance of winds in one unified scheme: 

%\begin{figure}[H]
%    \centering
%    \includegraphics[width=1\textwidth]{hopkins2008.jpg}
%    \caption{Cosmological AGN-Galaxy Co-Evolutionary model, as presented in \cite{hopkins2008cosmological}. in phases (d) (e) and (f) the AGN transition from an obscured state, in which accretion is also favored by the presence of large amounts of inflowing gas, to a phase where outflows develope to dissipate angular momentum or as a result of strong radiation pressure from high accretion rates, to an unobscured bright state. For more details on the whole model see the original source and references therein.} 
%    \label{fig:hop2008}
%\end{figure}

\begin{itemize}
    \item In 2001 the source was observed in a mild-Eddington state. Some amount of clumpy obscuring gas was present in the nuclear regions as it is currently accepted in AGN models, but was not significantly obscuring the source.
    \item In 2003 the source was obscured, due to gas intercepting the line of sight. This transition could either be explained in terms of a simple Keplerian orbit of an uneven, clumpy medium, but also as an inflowing motion of the gas. The 2005 observations favor the second option (see next bullet point).
    \item In 2005 we observed the AGN in a brighter state than before, accreting close to the Eddington limit. This may be due to the AGN accreting part of the gas that caused the increase in obscuration in 2003, with a resulting increase in intrinsic luminosity and a decrease in column density.
    \item In 2020 the source appeared to be obscured once again, but this time the absorbing material was ionized to a mild degree. The AGN dimmed back to its original state. We attribute the absorption to outflowing clouds intercepting the line of sight, pushed and ionized by the UFO, and ionized by the high radiation field generated from the previous accretion event.
    \item At the end of 2021 the source went back to its original mild-Eddington unobscured state, as the UFOs successfully cleared the surrounding gas. The AGN is now slightly brighter than it was in 2001, but not significantly. 

\end{itemize}

\begin{figure*}[h]
    \centering
   \includegraphics[width=\textwidth]{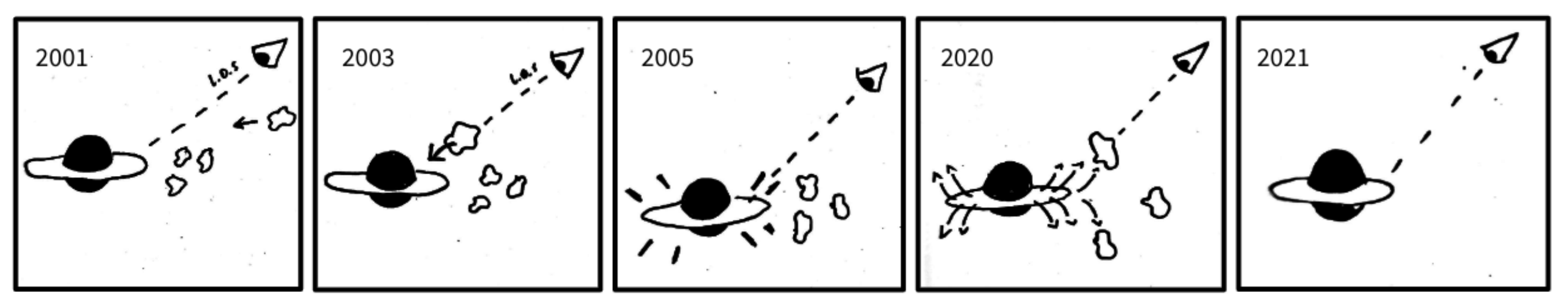}
    \caption{Cartoon of the proposed variability scenario (not to scale). Note that the episodic accretion is most likely happening in the form of a stream, rather than the single-cloud accretion event drawn in the simplistic figure.}
    \label{fig:toy}
\end{figure*}

Fig. \ref{fig:toy} shows a cartoon of the proposed scenario. 

Although we strongly support this model, based on our analysis, it must be noted that the timescales at which accretion rate changes propagate through standard Shakura-Sunyaev accretion disks are the viscous timescales:
\begin{equation}
\label{eq:tvisc}
    t_{vis} \sim 400 \bigg(\frac{H/R}{0.05}\bigg)^{-2} \bigg(\frac{\alpha}{0.03}\bigg)^{-1} \bigg(\frac{R}{150r_{g}}\bigg)^{3/2} M_{8} \ yrs,
\end{equation}
where $H/R$ measures the thickness of the disk, $\alpha$ is the viscous coefficient, $R$ the radius at which the timescale is computed, in units of gravitational radii, and $M_{8}$ is the SMBH mass in units of $10^{8} M_{\odot}$. Assuming standard prescriptions the viscous timescales for 2M0918 are of the order of some tens of years. This is a factor $\sim10$ times longer than the observed variability of 2 years, which would make our observations incompatible with theoretical predictions. However, it must be noted that the standard $\alpha$-disk has been recognized by theorists as an oversimplification for quite some time, and some solutions, such as a thick(er) disk or the role of magnetic fields, have been proposed to release assumptions, which would also shrink the timescales for accretion rate changes (See \citealp{lawrence2018quasar} for a discussion; see also \citealp{komossa2022extreme} and references therein). 

We also note that the episodic accretion event does not necessarily have to be communicated to the whole disk in order for luminosity to increase. In the case of tidal disruption events (TDEs), which are events in which Black Holes accrete stars passing in their vicinity, super-Eddington accretion disks can form on timescales of days (See e.g. \citealp{gezari2021tidal} for a review).
Although gas accretion and TDEs are different phenomena, we believe that variations in $\lambda_{EDD}$ can in principle happen on timescales shorter than the ones predicted by Eq. \ref{eq:tvisc} even for gas streams. For example, this could be the case if the accretion stream was not coplanar with the accretion disk (e.g. \citealp{chan2019tidal}).

\subsection{Future Perspectives}

We believe monitoring the source with simultaneous X-ray and Optical observations will help further constrain the nature of the variability of 2M0918. The future X-ray mission Einstein Probe (\citealp{yuan2022einstein}) dedicated to the monitoring of transients will provide further constraints on the variability of 2M0918. Furthermore, optical IFU observations will allow us to directly infer the geometry of the outflows, tightening our constraints on the ionized outflow energetics. This approach has been extensively used both at low-z (e.g \textcolor{black}{\citealp{venturi2021magnum, speranza2022warm, venturi2023complex, speranza2024multiphase}}) and at high-z (e.g. XID2028, \citealp{cresci2015blowin, cresci2023bubbles}). The
use of mm/sub-mm facilities, such as the Atacama Large Millimeter Array (ALMA), will allow us to probe the molecular phase of the outflow, which remains as of now totally unprobed and is thought to contain most of the gas involved in the outflow.

\begin{acknowledgements}
The authors thank Claudio Ricci, Giulia Tozzi, and Francesca Bonanomi for their contribution to data visualization and Nicola Locatelli for the useful discussion. 
This work is based on data from eROSITA, the soft X-ray instrument aboard \textit{SRG}, a joint Russian-German science mission supported by the Russian Space Agency (Roskosmos), in the interests of the Russian Academy of Sciences represented by its Space Research Institute (IKI), and the Deutsches Zentrum f\"ur Luft- und Raumfahrt (DLR). The \textit{SRG} spacecraft was built by Lavochkin Association (NPOL) and its subcontractors, and is operated by NPOL with support from the Max-Planck Institute for Extraterrestrial Physics (MPE).
The development and construction of the eROSITA X-ray instrument was led by MPE, with contributions from the Dr. Karl Remeis Observatory Bamberg \& ECAP (FAU Erlangen-Nuernberg), the University of Hamburg Observatory, the Leibniz Institute for Astrophysics Potsdam (AIP), and the Institute for Astronomy and Astrophysics of the University of T\"ubingen, with the support of DLR and the Max Planck Society. The Argelander Institute for Astronomy of the University of Bonn and the Ludwig Maximilians Universit\"at Munich also participated in the science preparation for eROSITA.
The eROSITA data shown here were processed using the eSASS/NRTA software system developed by the German eROSITA consortium.

Funding for the Sloan Digital Sky Survey (SDSS) has been provided by the Alfred P. Sloan Foundation, the Participating Institutions, the National Aeronautics and Space Administration, the National Science Foundation, the US Department of Energy, the Japanese Monbukagakusho, and the Max Planck Society. The SDSS Web site is http://www.sdss.org/. The SDSS is managed by the Astrophysical Research Consortium (ARC) for the Participating Institutions. The Participating Institutions are The University of Chicago, Fermilab, the Institute for Advanced Study, the Japan Participation Group, The Johns Hopkins University, Los Alamos National Laboratory, the Max-Planck-Institute for Astronomy (MPIA), the Max-Planck-Institute for Astrophysics (MPA), New Mexico State University, University of Pittsburgh, Princeton University, the United States Naval Observatory, and the University of Washington.

GL, MB, GM, and many Italian co-authors acknowledge support and funding from Accordo Attuativo ASI-INAF n. 2017-14-H.0 and from PRINMIUR2017PH3WAT  (‘Black hole winds and the baryon life cycle of galaxies’). 

IEL and BM received funding from the European Union’s
Horizon 2020 research and innovation program under
Marie Sklodowska-Curie grant agreement No. 860744
”Big Data Applications for Black Hole Evolution Studies” (BID4BEST).

R.A. received support for this work by NASA through the NASA Einstein Fellowship grant No HF2-51499 awarded by the Space Telescope Science Institute, which is operated by the Association of Universities for Research in Astronomy, Inc., for NASA, under contract NAS5-26555.

MP acknowledges grant PID2021-127718NB-I00 funded by the Spanish Ministry of Science and Innovation/State Agency of Research (MICIN/AEI/ 10.13039/501100011033).

GC, MB, MP, and EB acknowledge the support of the INAF Large Grant 2022 "The metal circle: a new sharp view of the baryon cycle up to Cosmic Dawn with the latest generation IFU facilities"

ZI acknowledges the support by the Excellence Cluster ORIGINS which is funded by the Deutsche Forschungsgemeinschaft (DFG, German Research Foundation) under Germany´s Excellence Strategy – EXC-2094 – 390783311.

\end{acknowledgements}

\bibliographystyle{aa} % style aa.bst
\bibliography{2MASS.bib} % your references Yourfile.bib
\begin{appendix}
\section{SED fitting with X-CIGALE}
\label{app:sed}
\begin{figure}[b]
    \centering
    \includegraphics[width=1.\columnwidth]{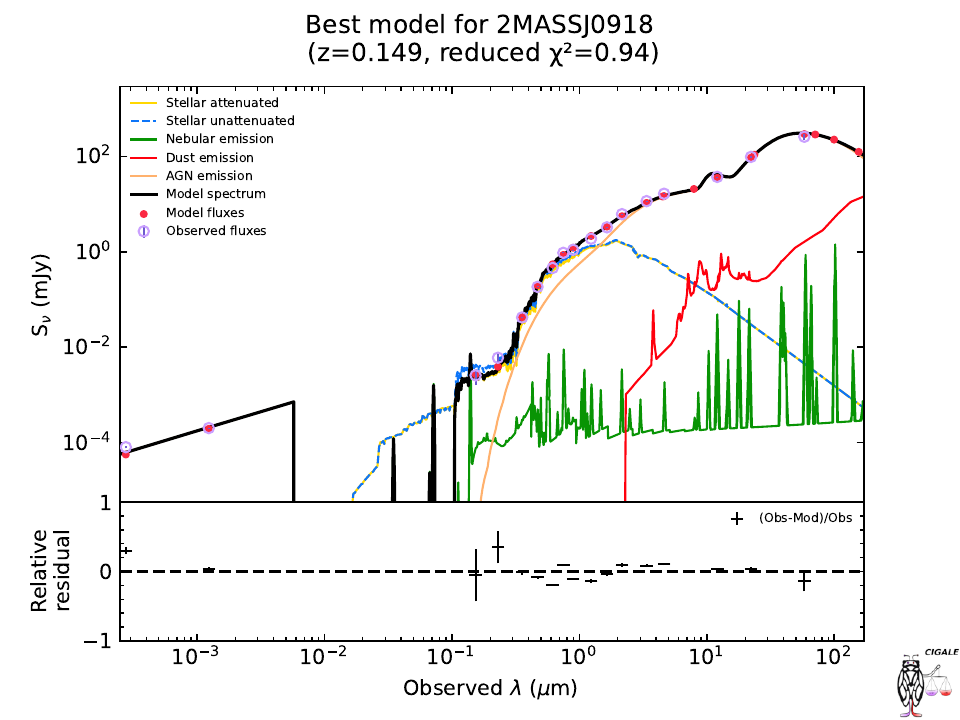}
    \caption{SED of 2M0918 as modeled with CIGALE, with residuals. As shown by the labels, the model includes templates for stellar, nebular, dust, and AGN emission, as well as dust attenuation.}
    \label{fig:sedkatia}
\end{figure}

In our analysis of 2MASS 0918+2117, we meticulously estimated crucial physical parameters through Spectral Energy Distribution (SED) fitting. The photometric data, obtained from the NED database, spanned diverse wavelength bands, including 2-10 keV, 0.5-2 keV, FUV, and NUV (GALEX), ugriz from SDSS, JHKs from 2MASS, W1-4 from WISE, and 24 and 160 microns from MIPS. We incorporated intrinsic flux data derived from the 2005 \textit{XMM-Newton} observation. Utilizing a Chabrier (\citealp{chabrier2003galactic}) initial stellar mass function (IMF) with solar metallicity and adopting a delayed star formation history for the stellar population, we employed Cigale models (\citealp{boquien2019cigale,yang2022fitting}) for the SED fitting. Dust extinction was accounted for using the model by \cite{dale2014two}. AGN contributions were delineated using Skirtor, allowing various inclinations, polar obscuration, and an extensive AGN fraction grid. The resulting best-fit, depicted in Fig. \ref{fig:sedkatia}, exhibited a reduced chi-square of 0.94.

The Bayesian posterior analysis yielded key parameters, notably an accretion luminosity ($L_\mathrm{acc} = 2.32 \pm 0.17 \times 10^{45}$ erg/s), a total bolometric luminosity ($L_\mathrm{bol} = 2.95 \pm 0.15 \times 10^{45}$ erg/s), and a host galaxy mass ($M_\star = 6.62 \pm 1.18 \times 10^{10}$ M$_\odot$). These findings offer valuable insights into the properties of 2MASS 0918+2117, particularly concerning the Bolometric Luminosity, which was directly calculated from the SED rather than relying on proxies, enhancing the precision and reliability of our results.

\end{appendix}

\end{document}